\documentclass[iop]{emulateapj}

\usepackage{amsmath, amsthm, amssymb, slashed}
\usepackage{color}
\usepackage[toc]{appendix}

\newcommand{\colvec}[2]{\left(\begin{array}{#1} #2 \end{array}\right)}

\newcommand{\p}{^{\prime}}
\newcommand{\dd}{{\rm d}}
\newcommand{\eps}{\varepsilon}

\usepackage{graphicx}

\begin{document}


\title{
  Toward more realistic analytic models of the heliotail: \\
  Incorporating magnetic flattening via distortion flows
}

\author{Jens Kleimann}
\affil{Ruhr-Universit\"at Bochum, Fakult\"at f\"ur Physik und Astronomie,
       Institut f\"ur Theoretische Physik IV, Bochum, Germany}
\email{jk@tp4.rub.de}

\author{Christian R\"oken} 
\affil{Universit\"at Regensburg, Fakult\"at f\"ur Mathematik, 
       Regensburg, Germany}  
\email{christian.roeken@mathematik.uni-regensburg.de}

\author{Horst Fichtner}
\affil{Ruhr-Universit\"at Bochum, Fakult\"at f\"ur Physik und Astronomie,
       Institut f\"ur Theoretische Physik IV, Bochum, Germany}
\email{hf@tp4.rub.de}

\author{Jacob Heerikhuisen}
\affil{
Department of Space Science and Center for Space Plasma and Aeronomic
Research, \\ University of Alabama in Huntsville, Huntsville, AL 35899, USA} 
\email{jacob.heerikhuisen@uah.edu}

\begin{abstract}
  Both physical arguments and simulations of the global heliosphere indicate
  that the tailward heliopause is flattened considerably in the direction
  perpendicular to both the incoming flow and the large-scale interstellar
  magnetic field. Despite this fact, all of the existing global analytical
  models of the outer heliosheath's magnetic field assume a circular cross
  section of the heliotail.
  To eliminate this inconsistency, we introduce a mathematical procedure by
  which any analytically or numerically given magnetic field can be deformed
  in such a way that the cross sections along the heliotail axis attain
  freely prescribed, spatially dependent values for their total area and
  aspect ratio. The distorting transformation of this method honors both the
  solenoidality condition and the stationary induction equation with respect
  to an accompanying flow field, provided that both constraints were already
  satisfied for the original magnetic and flow fields prior to the
  transformation. In order to obtain realistic values for the above
  parameters, we present the first quantitative analysis of the heliotail's
  overall distortion as seen in state-of-the-art three-dimensional hybrid
  MHD--kinetic simulations. \\
\end{abstract}


\section{Introduction}

Our knowledge about the large-scale structure of the heliosphere has recently
been increased significantly with the {\em Voyager}~1 and 2 spacecraft that
are by now exploring in~situ the local interstellar medium
\citep[LISM;][]{Gurnett-etal-2013} and the inner heliosheath
\citep[see, e.g.][]{Richardson-Burlaga-2013}, respectively. This progress is
accompanied by insights gained from the remote measurements of energetic
neutral atoms (ENAs) with the {\em Interstellar Boundary Explorer (IBEX)};
see the recent review by \citet{McComas-etal-2014b}. The {\em IBEX}
observations supplement those made with the {\em Voyagers} particularly
regarding the global structure of the heliosphere because they are not
limited to its upwind hemisphere but comprise its entirety. While in this
way the expected principal upwind--downwind asymmetry of the heliosphere
could be confirmed, the actual structure of its downwind hemisphere is still
under debate. However, there can be no doubt that the dominant feature of the
downwind heliosphere is the so-called heliotail, along which the solar wind
(SW) plasma eventually merges into the LISM. Consequently, this heliotail
--- predicted to exist already by \citet{Parker-1961} --- is a natural
feature of all classical models describing the interaction of the SW with the
LISM. Nonetheless, its detailed structure has been the subject of only a few
studies. In an early paper, \citet{Yu-1974} analyzed the magnetized
``wake'' of the SW under the influence of charge exchange with interstellar
hydrogen atoms. This line of research was continued only much later by
\citet{Jaeger-Fahr-1998} and \citet{Izmodenov-Alexashov-2003}. While both
papers confirmed the significance of this process, the latter authors
discussed a length of the  heliotail of several tens of thousands AU, whereas
the former favored a value of not more than 1,500~AU. Additionally,
\citet{Jaeger-Fahr-1998} recognized the potential significance of the
heliotail for the production of pick-up ions (PUIs). The related ENAs were
studied by \citet{Czechowski-Grzedzielski-1998}, who furthermore demonstrated
how the direction and strength of an interstellar magnetic field (ISMF) can be
deduced from the direction of the heliotail, an idea that was recently
revived by \citet{McComas-etal-2013}.
Quantitative models of the form and deflection of the heliotail in the
presence of a magnetized LISM were first presented by
\citet{Banaszkiewicz-Ratkiewicz-1989}, \citet{Matsuda-Fujimoto-1993}, and
\citet{Pogorelov-Matsuda-1998}.
\citet{Ratkiewicz_EA-2000}, while not explicitly addressing the heliotail
topic, provided a classification of the asymmetries and distortions resulting
from different ISMF orientations and mentioned a flattening effect, which
is also of relevance for the heliotail. Very recent studies looked at the
stability of the heliotail \citep{Pogorelov-etal-2014, Pogorelov_EA-2015} and
at the idea of its ``splitting'' \citep{Drake_EA-2015, Opher_EA-2015}, and,
thereby, revived work on two topics that were also discussed to some extent
already by \citet{Yu-1974}.

The heliotail is not only of significance for the flux of PUIs and ENAs but
also for that of cosmic rays (CRs). At the end of the last century, a
so-called heliomagnetotail anisotropy in the CR flux in the low-TeV range
was discovered by \citet{Nagashima-etal-1998}. This dipole-structured
anisotropy and its --- at least approximate --- relation to the heliotail
was confirmed by the latest generation of large-area detectors like MILAGRO,
the Tibet Air Shower, IceCube, and others \citep{Guillian-etal-2007,
Amenomori-etal-2010, Karapetyan-2010, Grigat-etal-2011}, showing an
enhancement in the permille range of the low-TeV particle flux in the
direction of the heliotail. A temporary denial of a physical
link between the heliotail and this anisotropy was based on the facts
(i) that the density in a gravitationally focused tail of interstellar
material would be much too low to explain the signal in terms of secondary
neutron production \citep{Drury-Aharonian-2008}, and (ii) that the
gyro-radii of particles at 10~TeV are about equal to the size of the
heliosphere or greater \citep{Abbasi-etal-2010}. Recently, however,
\citet{Lazarian-Desiati-2010} have proposed a new physical link between the
tail and the CR anisotropy by invoking magnetic reconnection as a process
that accelerates CRs in the 50~GeV to 10~TeV range in the heliotail.
While the existence of such a local source of CRs might be doubtful, in a
subsequent paper, \citet{Desiati-Lazarian-2013} considered the more likely
anisotropy-inducing effect of the ISMF, whose homogeneity on the scale of
the heliosphere is disturbed by the presence of the latter
\citep[see][and references therein]{Roeken_EA-2015}. They claim that the
large-scale CR anisotropy below 100~TeV is mostly shaped by particle
interactions with turbulent ripples generated by the interaction of the
heliospheric and the interstellar magnetic fields.

None of these explanations or first ``exploratory'' modeling attempts were
based on sophisticated heliospheric or advanced CR transport models,
which are, however, both necessary in order to derive quantitative results
that can be compared to observations. This gap has recently been fillede with 
the work by \citet{Schwadron_EA-2014} and \citet{Zhang-etal-2014}, who
studied the problem in much more detail by particularly computing CR
trajectories \citep[as was first done by][]{Washimi-etal-1999} rather than
by employing the CR diffusion approximation, which would represent a
conceptual extreme in that case. These two papers arrived at the conclusion
that a heliospheric impact on the CR anisotropy up to the TeV range must
indeed be expected.
Both studies, however, still have their drawbacks: Either an analytical model
that is probably too simplifying \citep{Schwadron_EA-2014}, or a numerical
input, which is difficult to handle \citep{Zhang-etal-2014}, is used for the
local ISMF configuration. For further analysis of the CR anisotropy problem,
it is therefore desirable to have a significantly improved analytical model
that incorporates crucial features of the ISMF.

A first step to construct such a model was made by \citet{Roeken_EA-2015},
who derived an exact analytical solution for the local ISMF assumed to be
frozen into the interstellar plasma flow. Not only this but, to the best of
our knowledge, all other analytical models of the local ISMF 
\citep{Whang-2010, Schwadron_EA-2014, Isenberg_EA-2015} feature a circular
cross section of the heliotail, i.e.\ far downtail, the heliopause takes the
shape of a semi-infinite cylinder. Nonetheless, the {\em IBEX} measurements
\citep{McComas-etal-2013} as well as several numerical studies
\citep[e.g.][]{Heerikhuisen_EA-2014, Wood_EA-2014} have confirmed the
intuitive notion that with growing heliocentric distance, the heliotail
becomes increasingly compressed perpendicular to the directions of the
undisturbed ISMF and the incoming interstellar flow.
Therefore, we propose a method by which both the magnetic field ${\bf B}$ and
a possibly associated flow field ${\bf u}$, or in fact any other such vector
field, may be deformed in a well-defined manner into a realistic
configuration, such that it is still maintaining both the field line topology
and the basic condition that the solenoidal magnetic field is frozen into
the (possibly, but not necessarily equally solenoidal) plasma flow. In other
words, if
\begin{equation}
  \label{eq:divb-0}
  \nabla \cdot {\bf B} = 0
\end{equation}
and
\begin{equation}
  \label{eq:induct}
  \nabla \times \left({\bf u} \times {\bf B} \right) = {\bf 0}
\end{equation}
were satisfied by the model's initial fields, then they will continue to be
satisfied by the deformed vector fields.

The outline of the paper is as follows. In Section~\ref{sec:deriv}, the
mathematical concept of distortion flows for the deformation of vector
fields is introduced and the equations relevant for the procedure are
derived. Section~\ref{sec:fitting} provides some guidance for the choice of
realistic parameters by extracting the relevant data from self-consistent
global simulations of the heliosphere, and applies them to the
\citet{Roeken_EA-2015} model. Finally, Section~\ref{sec:summary} completes
the paper with a summary and conclusions. The fact that our method honors
both the magnetic solenoidality condition (\ref{eq:divb-0}) and the
stationary induction equation (\ref{eq:induct}) is proven rigorously in
Appendix~\ref{app:cunning_proof}, while Appendices~\ref{app:wedge} and
\ref{app:fit-ellipse} provide some details on the general distortion
transformation and the employed data fitting method, respectively.\\

\section{Derivation of the Distortion Flow Method}
\label{sec:deriv}

In order to deform a given pair of magnetic and velocity fields,
we introduce the auxiliary field, ${\bf w}_0$, in which these fields are
advected. The geometrical shapes of the distorted fields can be controlled
by choosing a suitable form of ${\bf w}_0$. \\

\subsection{The Case of Constant Cross-sectional Areas}
\label{sec:deriv-const}

\begin{center}
  \begin{figure}[t]
    \begin{center}
      \includegraphics[width=0.4\textwidth]{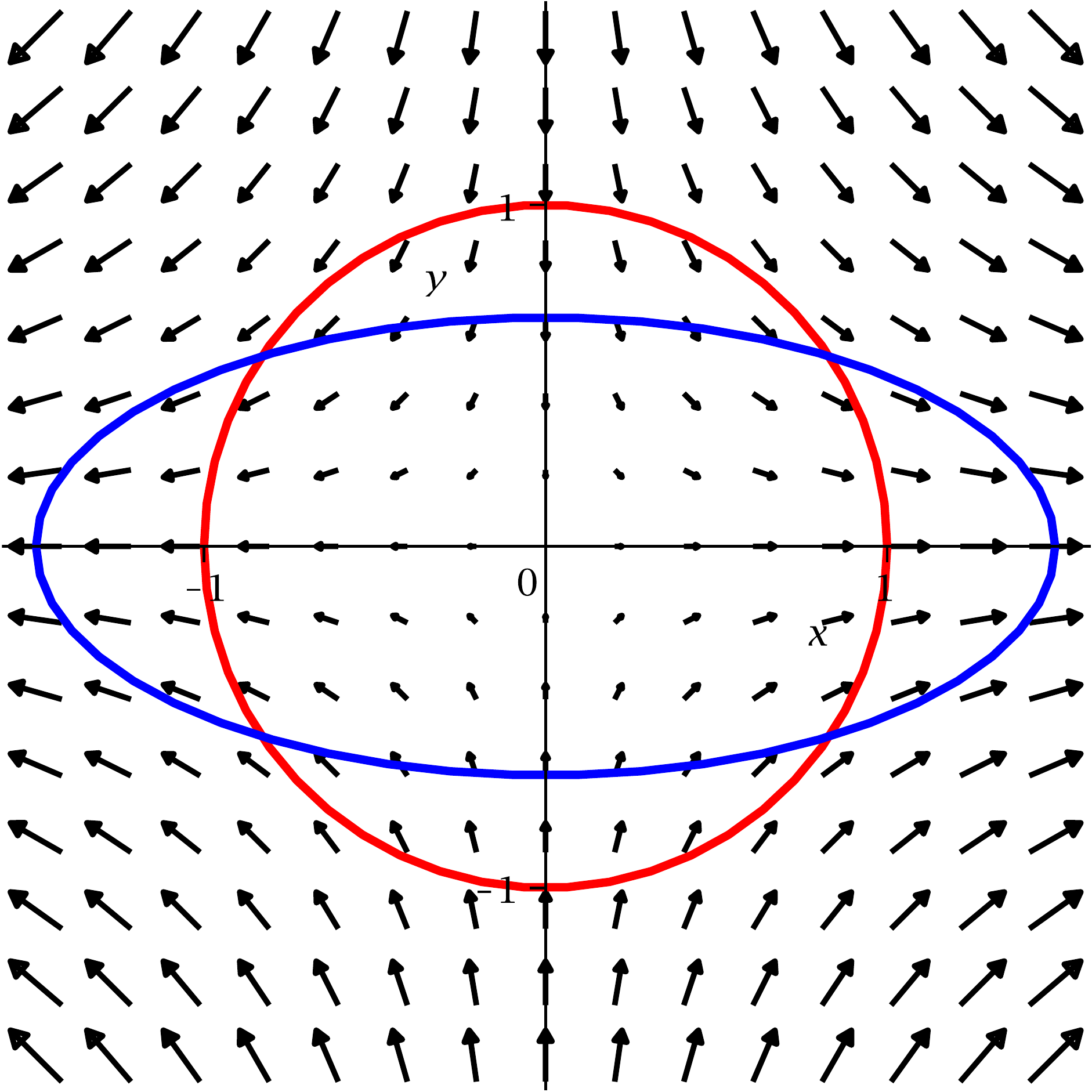}
    \end{center}
    \caption{\label{fig:distort-flow}
      The flow field (\ref{eq:w0}), together with an initially
      circular set of points at time $t=0$ (red) and the advected points at
      later time $t=t_1=0.2/\alpha>0$ (blue), when the distortion has
      reached an aspect ratio of $\eta=\exp(0.4) \approx 1.49$ according to
      Eq.~(\ref{eq:eta}). \\
    }
  \end{figure}
\end{center}

In view of the intended application to the heliotail, we want to contract
the initially circular cross-section in one direction and expand it in the
perpendicular direction. To this end, we fist consider the two-dimensional
(2D) velocity field
\begin{equation}
  \label{eq:w0}
  {\bf w}_0: \mathbb{R}^2 \rightarrow \mathbb{R}^2, \
  \colvec{c}{ x \\ y } \mapsto \colvec{r}{ \alpha \, x \\ -\alpha \, y}
\end{equation}
with constant $\alpha>0$. The equation of motion for an inertialess particle
being passively advected in this flow is
\begin{equation}
  \label{eq:motion}
  \dot{\bf r}(t) = {\bf w}_0[{\bf r}(t)] \ ,
\end{equation}
the solution of which, subjected to the initial condition
${\bf r}(0)={\bf r}_0$, reads
\begin{equation}
  \label{eq:xysol}
  \colvec{c}{ x(t) \\ y(t) } =
  \colvec{l}{ x_0 \exp(\alpha t) \\ y_0 \exp(-\alpha t) } \ .
\end{equation}
Specifically, an ensemble of particles which at $t=0$ forms a unit circle
will at $t=t_1 \ge 0$ have been deformed into an ellipse of half-axes
\begin{equation}
  \label{eq:ab}
  (a,b) := \big( \exp(\alpha t_1), \exp(-\alpha t_1) \big)
\end{equation}
with aspect ratio
\begin{equation}
  \label{eq:eta}
  \eta := \frac{a}{b} = \exp(2 \alpha t_1) \ge 1
\end{equation}
while maintaining its area ($\propto a \, b = 1$), in agreement with
$\nabla\cdot {\bf w}_0=0$ (see Fig.~\ref{fig:distort-flow}).

The idea is now to prescribe for each position $z$ along the heliotail axis
a ``target aspect ratio'' $\eta(z) = a(z)/b(z)$ into which the initially
circular cross sections are to be deformed by an embedding of (\ref{eq:xysol})
into three-dimensional (3D) space
\begin{eqnarray}
  x(t_1) &=& x_0 \exp \left[ +\alpha(z_0) \, t_1 \right] \\
  y(t_1) &=& y_0 \exp \left[ -\alpha(z_0) \, t_1 \right] \\
  z(t_1) &=& z_0
\end{eqnarray}
as the solution to Eq.~(\ref{eq:motion}), evaluated at a fixed $t=t_1$.
This motivates a mapping
$\mathbb{R}^3 \rightarrow \mathbb{R}^3$
\begin{eqnarray}
  \label{eq:xtr}
  x &=& a(z) \, x_0 \\
  y &=& b(z) \, y_0 \\
  z &=& z_0
\end{eqnarray}
from arbitrary undeformed coordinates ${\bf r}_0 = (x_0,y_0,z_0)$ at $t=0$
to deformed coordinates ${\bf r} = (x,y,z)$ at $t=t_1$, where $a(z)$ and
$b(z)$ are given by Eq.~(\ref{eq:ab}) with $\alpha=\alpha(z)$.
Note that this deformation has to be understood in the sense of a discrete
transformation $t = 0 \rightarrow t_1$, in contrast to the $t$-dependent,
continuous case (cf.\ Section~\ref{sec:relation}). Moreover, the requirement
to ensure the constancy of the cross-sectional area $\pi \, a(z) \, b(z)$
for any $z$ implies
\begin{equation}
  \label{eq:ab-from-eta}
  a(z) = \eta(z)^{1/2} \quad \mbox{and} \quad
  b(z) = \eta(z)^{-1/2} \ .
\end{equation}

To see how a deformed vector ${\bf P} \in \{ {\bf u}, {\bf B} \}$ is obtained
via the distortion field ${\bf w}_0$, consider the unsqueezed vector
${\bf P}_0$ located at ${\bf r}_0$ and pointing in the direction of
${\bf \delta r}_0$, i.e.\ ${\bf P}_0 = \nu \, {\bf \delta r}_0$, where
${\bf \delta r}_0$ is a small displacement vector, which may also be pictured
as a small segment of a field line. The constant $\nu$ is introduced to
warrant dimensional consistency, and can be used to accommodate scalings of a
desired magnitude. Via Eq.~(\ref{eq:xtr}), the $x$ component of the deformed
vector ${\bf P}$ at its new position ${\bf r}$ becomes
\begin{equation}
  \begin{split}
    P_x ({\bf r})
    &= \nu \, [ (x + \delta x) - x] \\
    &= \nu \, [ a(z+\delta z) \ (x_0+\delta x_0) - a(z) \ x_0 ] \\
    &\approx \nu \, [ a(z) \ \delta x_0 + a\p(z) \, x_0 \, \delta z ] \\
    &= a(z) \, P_{0x}({\bf r}_0) + a\p(z) \, x_0 \, P_{0z}({\bf r}_0) \ ,
    \label{eq:Pxo2Px}
  \end{split}
\end{equation}
where a prime denotes differentiation with respect to $z$, and the
``$\approx$'' symbol has been used only because the terms of
${\cal O}(\delta^2)$ have been neglected in the Taylor expansion of
$a(z+\delta z)$.
The $y$ component is found in complete analogy, and since the $z$
component remains unaffected (i.e.\ $z_0=z$), we end up with
\begin{equation}
  \colvec{c}{ P_x \\ P_y \\ P_z } =
  \colvec{c}{ a \, P_{0x} \\ b \, P_{0y} \\ 0 } +
  P_{0z} \colvec{c}{ (a\p / a) \, x \\ (b\p / b) \, y \\ 1 } \ .
  \label{eq:Pxo2Px_vec}
\end{equation}
In other words, given an arbitrary, unsqueezed vector field ${\bf P}_0$,
together with the function $\eta(z)$ specifying the desired aspect ratio
relative to a circle, the components of the distorted field ${\bf P}$ at
position ${\bf r}$ can be obtained by
\begin{enumerate}
\item finding the corresponding ``starting location''
  \mbox{${\bf r}_0({\bf r}) = \big(x/a(z), y/b(z), z \big)$};
\item evaluating ${\bf P}_0({\bf r}_0)$ at this position; and
\item transforming the result according to Eq.~(\ref{eq:Pxo2Px_vec}).
\end{enumerate}
For scalar quantities $Q$ such as mass density or temperature, which contain
no directional information, the proper form after distortion is simply
$Q({\bf r}) = Q_0({\bf r}_0)$.

\begin{widetext}
  \begin{center}
    \begin{figure}[t]
      \includegraphics[width=\textwidth]{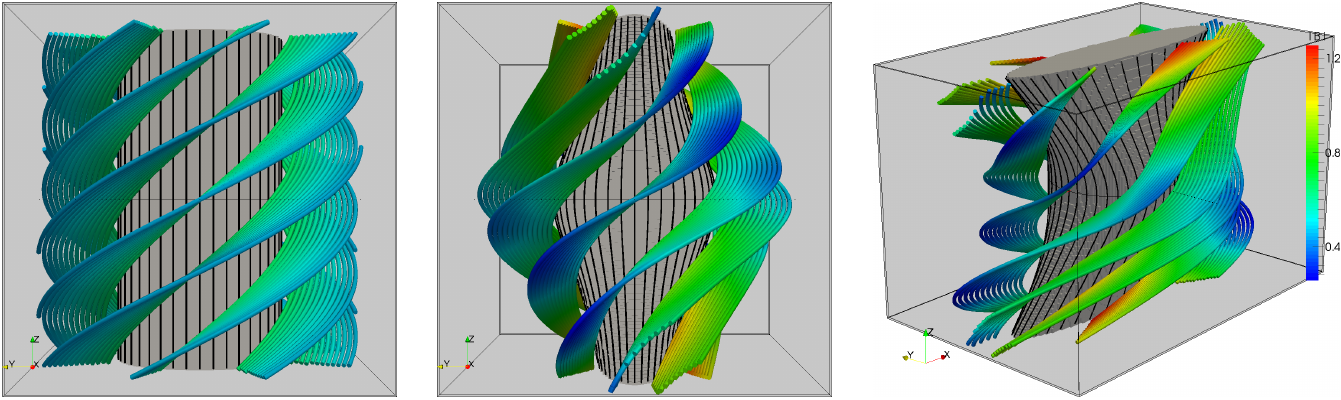}
      \caption{\label{fig:fluxtube}
        Left: selected field lines of the undistorted flux tube
        (\ref{eq:fluxtube-0}).
        Middle: the corresponding magnetic field lines (\ref{eq:fluxtube-1})
        after a distortion according to Eqs.~(\ref{eq:d-a})--(\ref{eq:d-b})
        has been applied, viewed along the major ($x$)~axis.
        Right: the same situation viewed from a different angle.
        The gray surface indicates the cylinder $\rho^2=x^2+y^2=1$ (left),
        which is transformed into the shape $[x/a(z)]^2+[y/b(z)^2]=1$
        via the squeezing transformation (\ref{eq:Pxo2Px_vec}).
      }
    \end{figure}
  \end{center}
\end{widetext}
In the above derivation, the distortion flow ${\bf w}_0$ (which mediates the
deformation) is an auxiliary field that must be carefully distinguished
from the actual velocity field ${\bf u}$ upon which the transformation acts.
${\bf w}_0$ is thus simply a mathematical object that acts similar to a
flow field, although no physical flow of matter is associated with it.
Likewise, the parameter $t$, which we have referred to as ``time'' for the
sake of clarity, is rather to be understood as a deformation parameter
controlling the extent of deformation from the original field
${\bf P}_0 \in \{ {\bf u}_0, {\bf B}_0 \}$ (at ``time'' $t=0$) to its
distorted counterpart ${\bf P} \in \{ {\bf u}, {\bf B} \}$ at $t=t_1$.
We continue to use the symbol $t$ since the stationary induction equation
(\ref{eq:induct}) implies that the configuration formed by ${\bf u}$ and
${\bf B}$ does not depend on physical time, such that no confusion should
arise from this terminology. \\

\subsection{A Simple Example}

To illustrate the procedure, consider the cylindrical magnetic flux tube
\begin{equation}
  \label{eq:fluxtube-0}
  \begin{split}
    {\bf B}_0({\bf r}_0) &= \frac{\rho_0}{1+\rho_0^2} \ {\bf e}_{\varphi_0}
    + \frac{1}{1+\rho_0^2} \ {\bf e}_{z_0} \\
    &= \frac{1}{1+x_0^2+y_0^2} \colvec{c}{ -y_0 \\ x_0 \\ 1 }
  \end{split}
\end{equation}
(where $\rho_0 := \sqrt{x_0^2+y_0^2}$ denotes the cylindrical radial
coordinate and $\varphi_0 := \arctan(y_0/x_0)$ the corresponding angular
coordinate), together with a helical ``swirl flow''
\begin{equation}
  \label{eq:swirl-0}
  {\bf u}_0({\bf r}_0) = \Omega \, \rho_0 \ {\bf e}_{\varphi_0}
  + V \ {\bf e}_{z_0} = \colvec{c}{ -\Omega \, y_0 \\ \Omega \, x_0 \\ V } \ ,
\end{equation}
characterized by constants $\Omega$, $V \in \mathbb{R}$. It can easily be
shown that these fields satisfy both
$\nabla_0 \times ({\bf u}_0 \times {\bf B}_0)={\bf 0}$ and
$\nabla_0 \cdot {\bf B}_0=0$.
In this example, we choose a deformation with a spatially variable aspect
ratio $\eta(z)= 1+z^2$, which implies
\begin{eqnarray}
  \label{eq:d-a}  a(z) &=&          \sqrt{1+z^2} \\
  \label{eq:d-b}  b(z) &=& \frac{1}{\sqrt{1+z^2}}
\end{eqnarray}
according to Eq.~(\ref{eq:ab-from-eta}).
Since we are interested in the explicit forms of the distorted fields
${\bf B}$ and ${\bf u}$, we now express all points $(x_0,y_0,z_0)$ in
${\bf P}_0({\bf r}_0)$ through $(x,y,z)$ and apply the transformation
(\ref{eq:Pxo2Px_vec}) to the fields (\ref{eq:fluxtube-0}) and
(\ref{eq:swirl-0}). Thus, with
\begin{eqnarray}
  P_x(x,y,z) &=& a(z) \
  P_{0x} \left( \frac{x}{a(z)}, \frac{y}{b(z)}, z \right) \nonumber \\
  && + \frac{a\p(z)}{a(z)} \, x \, P_{0z}
  \left( \frac{x}{a(z)}, \frac{y}{b(z)}, z \right) \\
  P_y(x,y,z) &=& b(z) \
  P_{0y} \left( \frac{x}{a(z)}, \frac{y}{b(z)}, z \right)\nonumber \\
  && + \frac{b\p(z)}{b(z)} \, y \, P_{0z}
  \left( \frac{x}{a(z)}, \frac{y}{b(z)}, z \right) \\
  P_z(x,y,z) &=&
  P_{0z} \left( \frac{x}{a(z)}, \frac{y}{b(z)}, z \right) \ ,
\end{eqnarray}
we obtain the distorted fields
\begin{eqnarray}
  \label{eq:fluxtube-1} \nonumber
  {\bf B} &=& \frac{1}{1+x^2+y^2(1+z^2)^2+z^2} \\
  &&\times
  \colvec{c}{xz-y (1+z^2)^2 \\ x-yz \\ 1+z^2} \\
  {\bf u} &=& \frac{1}{1+z^2}
  \colvec{c}{ x z \, V - y (1+z^2)^2 \, \Omega
    \\ x \, \Omega - y z \, V \\ (1+z^2) \, V} \ ,
\end{eqnarray}
which can again be verified to satisfy the desired constraints
(\ref{eq:divb-0}) and (\ref{eq:induct}). Fig.~\ref{fig:fluxtube} illustrates
the magnetic field configuration before and after the deformation. \\

\subsection{Relation to Inductive Flux Transport}
\label{sec:relation}

The advection of field lines via distortion flows is reminiscent of the
transport of frozen-in field lines via the classical induction equation of
ideal MHD \citep[e.g.][]{Childress-Gilbert_BOOK}. It is therefore reasonable
to ask how the method detailed above relates to physical flux transport via
an induction-type equation
\begin{equation}
  \label{eq:dPdt_induct}
  \partial_t {\bf P} = \nabla \times ( {\bf w} \times {\bf P})
\end{equation}
that would lead to a similar result, however, for continuous deformations
with respect to $t$ described by the transformation
$({\bf r}_0,t) \mapsto \big( {\bf r}({\bf r}_0, t), t \big)$.
To clarify this point, we first derive the explicit evolution equation
that is actually solved by the fields given by the distortion flow method.
In the spirit of Eq.~(\ref{eq:Pxo2Px}), but now with a varying
time/deformation parameter $t$, the vector field
${\bf P}({\bf r},t)$ at position ${\bf r} = {\bf r}({\bf r}_0,t)$ and
consecutive times $t$ and $t+\delta t$ is
\begin{eqnarray}
  \label{eq:P_to}
  {\bf P} \big({\bf r}, t \big) &=&
  \nu \, \big[ \hat{\bf r}({\bf r}, t) - {\bf r}({\bf r}_0, t) \big] =
  \nu \, \delta {\bf L}({\bf r},t) \\
  \label{eq:P_to+dt}
  {\bf P} \big( {\bf r}, t+\delta t \big) &=&
  \nu \, \big[ \hat{\bf r}({\bf r}, t+\delta t) - {\bf r}({\bf r}_0, t+\delta t) \big] \\
  &=& \nonumber  \nu \, \delta {\bf L}({\bf r},t+\delta t) \ ,
\end{eqnarray}
where $\delta {\bf L}$ denotes a small, time-dependent displacement vector,
and $\hat{\bf r} := {\bf r} + \delta {\bf L}$. Note that
$t \mapsto {\bf r}({\bf r}_0, t)$ is the trajectory starting at ${\bf r}_0$,
along which the ${\bf P}$ vector moves due to the action of the distortion
flow ${\bf w}$, yielding evolutionary deformation. Expanding
(\ref{eq:P_to+dt}) for small $\delta t$, subtracting (\ref{eq:P_to}), and
neglecting the terms of ${\cal O}(\delta t^2)$, we obtain
\begin{equation}
  \label{eq:my_beans}
  \begin{split}
    \left( \frac{\dd {\bf r}}{\dd t} \cdot {\nabla} \right) {\bf P}
    + \partial_t {\bf P}
    &= \nu \left( \frac{\dd \hat{\bf r}}{\dd t}  -
      \frac{\dd {\bf r}}{\dd t} \right) \\
    &= \nu \, \big[ {\bf w} ( {\bf r} + \delta {\bf L} )
    - {\bf w} ({\bf r}) \big] \\
    &\approx \, \nu \big( \delta {\bf L} \cdot \nabla \big)
    {\bf w} ({\bf r}) \ ,
  \end{split}
\end{equation}
where the evolution equation
\begin{equation}
  \label{eq:evolution}
  \frac{\dd {\bf r}}{\dd t} = {\bf w}({\bf r})
\end{equation}
and a Taylor expansion of ${\bf w} ({\bf r}+\delta {\bf L})$ for small
$\delta {\bf L}$ up to first order have been used. Substituting
Eq.~(\ref{eq:P_to}) and again (\ref{eq:evolution}) into (\ref{eq:my_beans}),
one finds
\begin{equation}
  \begin{split}
    \label{eq:dPdt_distort}
    \partial_t {\bf P} &= ({\bf P} \cdot \nabla) {\bf w}
    - ({\bf w} \cdot \nabla) {\bf P} \\
    &= \nabla \times ( {\bf w} \times {\bf P})
    - (\nabla \cdot {\bf P}) {\bf w}
    + (\nabla \cdot {\bf w}) {\bf P} \ ,
  \end{split}
\end{equation}
which shows explicitly that the distortion flow method becomes equivalent to
solving the ``induction equation'' (\ref{eq:dPdt_induct}) for the special
case in which both $\nabla \cdot {\bf w}$ and $\nabla \cdot {\bf P}$ vanish.
So in principle, one could as well use (\ref{eq:dPdt_induct}) to obtain some
form of distorted fields, in which case $\partial_t (\nabla \cdot {\bf P})=0$
would even be satisfied unconditionally, i.e.\ without requiring any
constraints on the value of $\nabla \cdot {\bf w}$.
However, as can be shown using an analysis similar to the one presented
toward the concluding part of Appendix~\ref{app:cunning_proof}, the
alternative evolution equation (\ref{eq:dPdt_induct}) can maintain
$\nabla \times ( {\bf u} \times {\bf B})={\bf 0}$ only for the special case
$\nabla \cdot {\bf P}=0=\nabla \cdot {\bf w}$, in which it becomes identical
to Eq.~(\ref{eq:dPdt_distort}) anyway. Additionally, analytical solutions to
(\ref{eq:dPdt_induct}) are usually difficult to obtain even for simple
choices of ${\bf w}$, whereas for the method proposed here, it suffices to
solve the equation of motion (\ref{eq:evolution}), which will be much
simpler to do in most cases. \\

\subsection{Generalization to Varying Cross-sectional Areas}
\label{sec:deriv-noco}

For a given application, the condition that cross sections can change their
shapes but not their absolute areas may not always be satisfied. (This
certainly holds true in the case of the heliospheric tail, as will be shown
in Section~\ref{sec:fit-csvari}.)
Therefore, we now wish to relax this constraint, such that $a(z)$ and $b(z)$
may be chosen independently of one another. Guided by our findings from the
simpler situation considered so far, we will again take the distortion to
be incompressible. This implies that the latter will no longer be confined
to separate $x$--$y$ planes, but that material from one $z$ layer may be
displaced into adjacent layers, i.e.\ the distortion flow corresponding to
that of Eq.~(\ref{eq:w0}) will attain a non-zero $z$~component.

So consider $a(z)$ and $b(z)$ as given independently of one another. In an
incompressible distortion, the spatial volume element is to be conserved,
i.e.
\begin{equation}
  \label{eq:const_dV}
  \dd x_0 \ \dd y_0 \ \dd z_0 = \dd x \ \dd y \ \dd z \ .
\end{equation}
Additionally, we now allow for a prescribed displacement $m(z)$ along the
$x$ axis as a third free parameter besides $a(z)$ and $b(z)$. The $x$ and
$y$ components of the transformation thus become
\begin{equation}
  \label{eq:xxpyyp}
  x_0 = \frac{x-m(z)}{a(z)} \quad \mbox{and} \quad
  y_0 = \frac{y}{b(z)} \ .
\end{equation}
A similar displacement in the $y$~direction could easily be accounted for as
well. We will refrain from doing so for simplicity of argument, and also
because it will not be relevant for the intended application to the heliotail
(see Section~\ref{sec:fitting}).  While various choices of $z_0=z_0(x,y,z)$
would be consistent with Eq.~(\ref{eq:const_dV}), we restrict ourselves to
the special case of $z_0=z_0(z)$, describing the situation that all particles
from the plane at $z_0$ are being shifted into another plane at $z$.
The general case of $z_0=z_0(x,y,z)$ is addressed in Appendix~\ref{app:wedge}.
Consequently, the $z$ component of the distortion transformation has to
satisfy the relation
\begin{equation}
  \label{eq:zzp}
  z_0 = \int_0^{z_0} \dd \tilde{z} \stackrel{!}{=}
  \int_0^z a(\tilde{z}) \, b(\tilde{z}) \ \dd \tilde{z} =: F(z)
\end{equation}
in order to remain consistent with Eq.~(\ref{eq:const_dV}).
As can be seen in this equation, $z=0$ is to be interpreted as the
``reference height'' at which the transformation induces no vertical
displacement of ${\bf P}$, i.e.\ the planes $z_0=0$ and $z=0$ coincide.

Repeating the derivation of Eq.~(\ref{eq:Pxo2Px}), we are led to
\begin{equation}
  \begin{split}
    P_x({\bf r}) =&\
    \nu \ [ (x+\delta x) - x ] \\
    =&\ \nu \ [ a(z+\delta z) \ (x_0+\delta x_0) + m(z+\delta z) \\
    &- a(z) \ x_0 - m(z) ] \\
    \approx&\ \nu \ [ a(z) \ \delta x_0
    + a\p(z) \ \delta z \ x_0
    + m\p(z) \ \delta z ] \\
    =&\ a(z) \ P_{0x} + [ a\p(z) \, x_0 + m\p(z) ] P_z \ .
    \label{eq:Px_deriv}
  \end{split}
\end{equation}
The corresponding equation for $P_y$ is found in complete analogy, except
that both $m(z)$ and $m\p(z)$ are absent. Introducing now $z=K(z_0)$ as the
inverse of $z_0=F(z)$, we obtain
\begin{equation}
  \begin{split}
    P_z({\bf r})
    =&\ \nu \, [ (z +\delta z) - z ] \\
    =&\ \nu \, [ K(z_0+\delta z_0) - K(z_0) ] \\
    =&\ \nu \, [ \partial_{z_0} K(z_0) \ \delta z_0 + {\cal O}(\delta z^2) ]
    \\ \approx&\ [a(z) \, b(z)]^{-1} P_{0z}
    \label{eq:Pz_deriv}
  \end{split}
\end{equation}
for the $z$ component.
Inserting (\ref{eq:Pz_deriv}) into (\ref{eq:Px_deriv}) and suppressing the
argument $z$ in $a$, $b$, and $m$, the complete vector transformation
becomes
\begin{equation}
  \label{eq:distort}
  \colvec{c}{ P_x \\ P_y \\ P_z } =
  \colvec{c}{ a \, P_{0x} \\ b \, P_{0y} \\ 0 } + \frac{P_{0z}}{a b}
  \colvec{c}{ a\p x_0 + m\p \\ b\p y_0 \\ 1 } \ .
\end{equation}
While in this form it looks very similar to its analog (\ref{eq:Pxo2Px_vec}),
the most important difference is hidden in the additional requirement to
obtain $z_0$ as a function of $z$, which involves the evaluation of the
integral (\ref{eq:zzp}). This procedure will be illustrated in
Section~\ref{sec:fit-csvari}. \\

\section{Realistic Parameters for the Heliotail}
\label{sec:fitting}

Returning to our intended application, the heliotail, the above
considerations naturally provoke the question of how the $z$ profiles for
the squeezing functions $a$, $b$ and the displacement $m$ should be chosen
in order to obtain a configuration that resembles actual heliospheric
conditions as closely as possible.

\subsection{Data Source and Fitting Method}

To provide some guidance for these choices, we turn to the 3D plasma-neutral
simulations of the interaction between the SW and the LISM performed by
\citet{Heerikhuisen_EA-2014}. These simulations employ an MHD solver
\citep{Pogorelov_EA-2006} for the ionized plasma and a Monte-Carlo
particle-based Boltzmann solver \citep{Heerikhuisen_EA-2006} for neutral
hydrogen. The neutral component is coupled to the plasma through
charge-exchange collisions that generate energy and momentum source terms
which are applied to the MHD equations \citep{Heerikhuisen_Pogorelov-2010}.
For the simulations considered here, the ion and neutral components were
iterated until a steady-state solution to the SW--LISM boundary value problem
was obtained.
An important aspect of these simulations is the removal of energy from the
SW through charge exchange with cold LISM hydrogen atoms as it flows down the
heliotail. This leads to a reduction in pressure, and hence a decrease in the
heliotail cross section at large distances. The simulations were performed on
a spherical grid where the polar axis is aligned with the solar rotation
axis, using an angular resolution of $3^{\circ}$ in azimuth, $1.5^{\circ}$ in
colatitude, and a non-uniform radial grid with 284 cells. The inner boundary
is located at 10~AU, and the outer boundary at 1,000~AU. The resulting
heliotail extends well beyond the outer boundary in the downstream LISM
direction. The cell size increases with radial distance, which can make it
difficult to accurately determine the heliopause location in the distant
heliotail. However, the bulk properties of the plasma and neutral
populations and the overall shape of the heliosphere are not significantly
affected by this reduction in resolution near the outer boundary.

Since these simulations were performed for magnetic field strengths of
$B_{\rm ism} \in \{1,2,3,4\}\times 0.1$~nT, they resulted in four distinct
3D configurations after a steady state had been reached. Here, these four
datasets were then subjected to the following analysis: For all combinations
$j$ of magnetic field strengths $B_{\rm ism}$ and chosen  $x$--$y$ coordinate
planes with $z \in \{-100,-200, \ldots, -800\}$~AU, the heliopause locations
${\cal H}_j$ were determined as the isocontours in the $x$--$y$ plane at
which the plasma temperature has increased by a factor of five over its
undisturbed LISM value at the outer computational boundary
\citep[see Fig.~6 in][]{Heerikhuisen_EA-2014}.
Each ${\cal H}_j$ is then typically given by $N \approx 200$ pairs of
coordinates $(x_i,y_i)_j$, $i = 1, \ldots, N$, which have been rotated
around the inflow ($z$) axis such that the $x$ direction coincides with the
direction of ${\bf B}_{\rm ism}$ at infinity when projected onto the $x$--$y$
plane. An ellipse
\begin{equation}
  {\cal E}_j:
  \left(\frac{x_i-m_j       }{a_j}\right)^2 +
  \left(\frac{y_i-m_{\perp,j}}{b_j}\right)^2 = 1
\end{equation}
(in which the second index $j$ of $x_i$ and $y_i$ has been suppressed),
with major semi-axis $a_j$, minor semi-axis $b_j$, and center coordinates
$(m_j, m_{\perp,j})$ was then fitted to each ${\cal H}_j$ contour by varying
these four parameters until the total ``difference area'' $\Delta_j$
(i.e.\ the area covered by the interiors of either ${\cal H}_j$ or
${\cal E}_j$, but not both) attained its global minimum.
The explicit form of $\Delta_j$ is given in Appendix~\ref{app:fit-ellipse}.
Here $a_j$, $b_j$, and $m_j$ play the same role as the functions $a$, $b$,
and $m$ in the above derivation, except that they are no longer dimensionless
but measured in absolute units of AU. 
\begin{figure}[b]
  \begin{center}
    \includegraphics[width=0.47\textwidth]{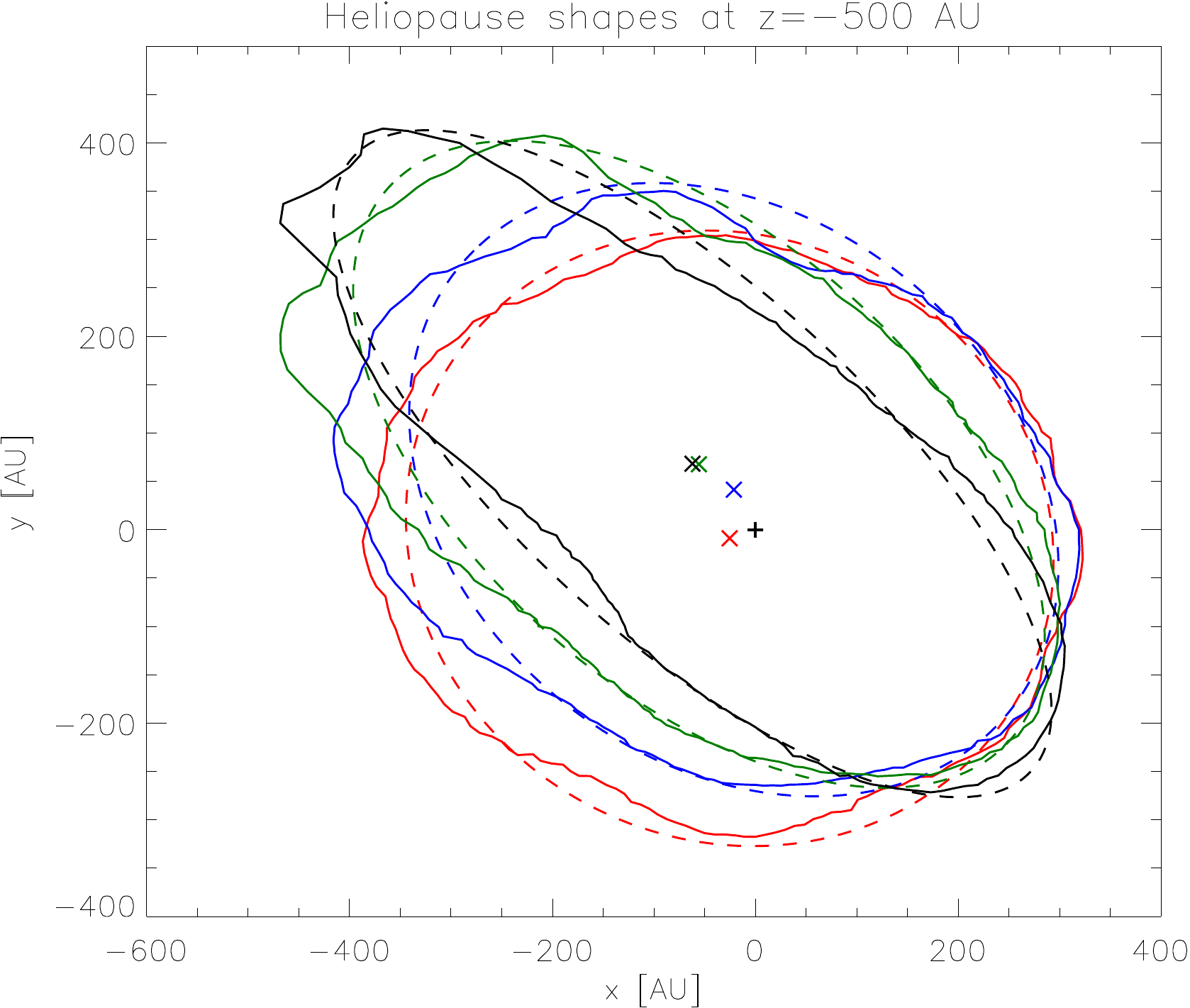}
  \end{center}
  \caption{\label{fig:helli_slice-z5}
    Heliotail contours at $z=-500$~AU as derived from the hybrid simulation
    for $B_{\rm ism}$ values of 0.1 (red), 0.2 (blue), 0.3 (green), and 0.4~nT
    (black), together with the respective fitting ellipses (dashed). The
    ``$+$'' symbol marks the inflow axis \mbox{($x=0=y$)}, while the
    ``$\times$'' symbols indicate the centers of the respective ellipses. \\
  }
\end{figure}
Note that in order to obtain sufficiently good fitting results, we found it
necessary also to allow for a displacement $m_{\perp,j}$ in the direction of
${\cal E}_j$'s minor axis as a fourth parameter.

To illustrate this process, Fig.~\ref{fig:helli_slice-z5} shows the fitting
results for the case of $z=-500$~AU, which was also used
\begin{figure}[h]
  \begin{center}
    \includegraphics[width=0.47\textwidth]{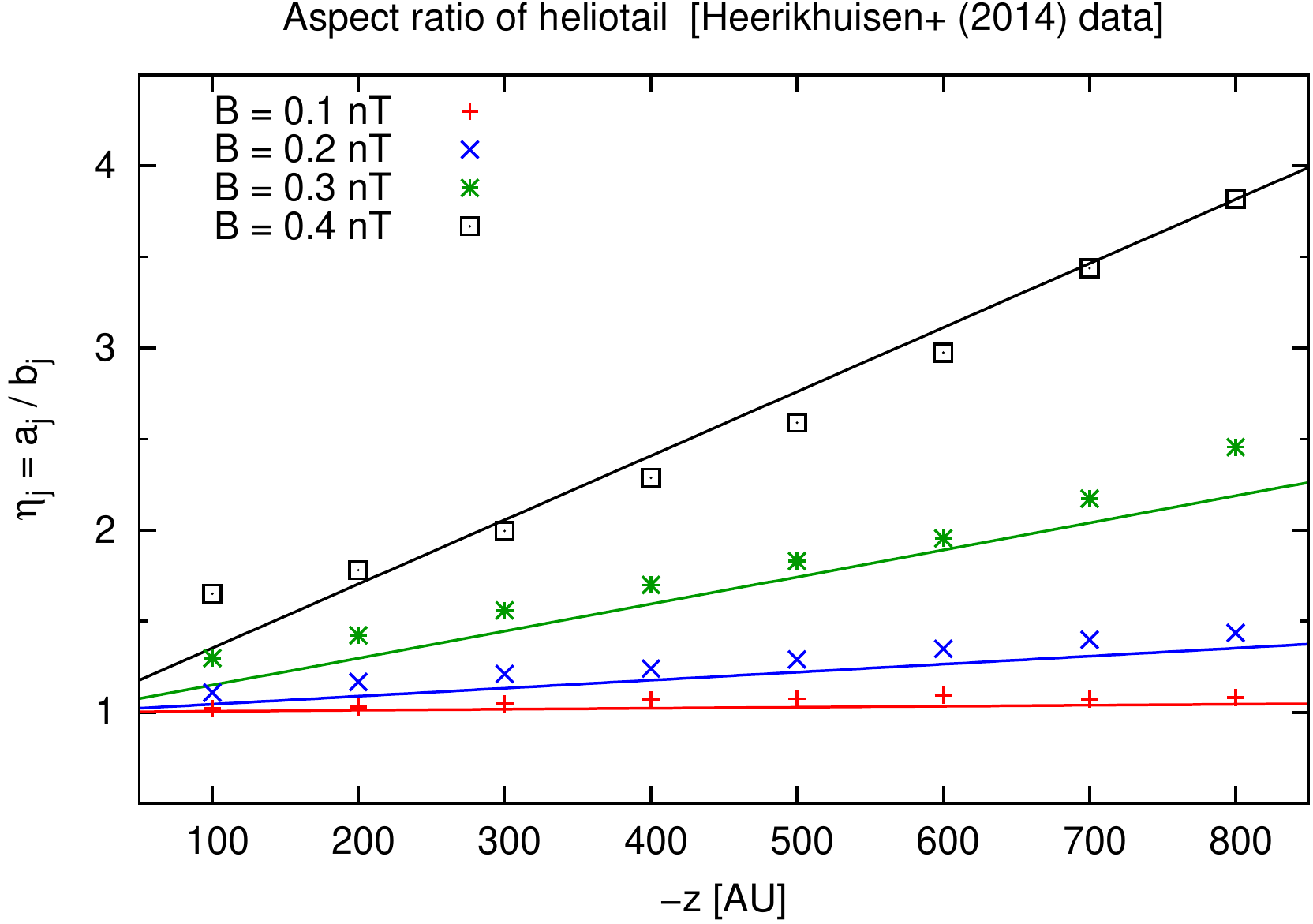}
  \end{center}
  \caption{\label{fig:helli_a-r}
    Aspect ratios $\eta_j$  of fitting ellipses, together with fitting
    function (\ref{eq:fit-eta}). \\
  }
\end{figure}
\begin{figure}[h]
  \begin{center}
    \includegraphics[width=0.47\textwidth]{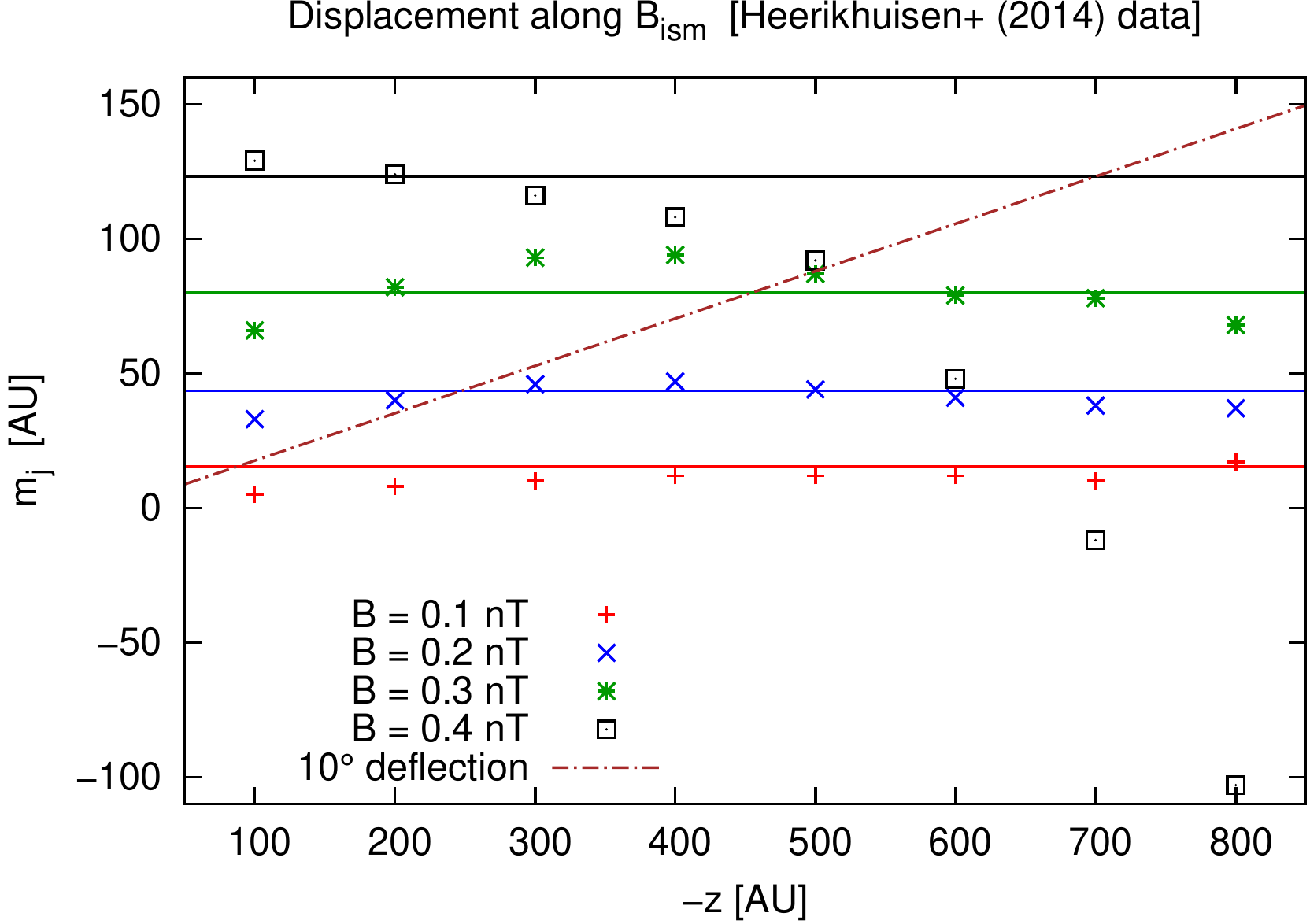}
  \end{center}
  \caption{\label{fig:helli_m-m}
    Linear displacement of the heliotail core axis against the inflow
    direction, together with fitting function
    (\ref{eq:fit-m}). The dashed line marks a hypothetical deflection of
    $10^{\circ}$ against the inflow axis. For the $B=0.4$~nT case, data
    points at $z<-400$~AU were disregarded when setting up the fitting
    relation (\ref{eq:fit-m}). This is justified by the observation that
    for these cases the heliopause contours develop a pronounced
    pear-shaped asymmetry that renders the elliptic approximation
    questionable. \\
  }
\end{figure}
for Fig.~6 in
\citet{Heerikhuisen_EA-2014}. The minus sign honors the convention that the
LISM flow is incident from $z \rightarrow +\infty$, which will be exploited
in Section~\ref{sec:app_helio}.
The parameters $\eta_j := a_j/b_j$ and $m_j$ thus derived are displayed as
functions of $z$ in Figs.~\ref{fig:helli_a-r} and \ref{fig:helli_m-m}
along with the heuristic fitting functions
\begin{eqnarray}
  \label{eq:fit-eta}
  \eta_{\rm fit}(B_{\rm ism}, z) &:=& 1 + 3.27
  \left( \frac{-z}{100~\mbox{AU}} \right)
  \left( \frac{B_{\rm ism}}{\mbox{nT}} \right)^{2.5} \\
  \label{eq:fit-m}
  m_{\rm fit}(B_{\rm ism}, z) &:=& \left( 487~\mbox{AU} \right)
  \left( \frac{B_{\rm ism}}{\mbox{nT}} \right)^{1.5} \ ,
\end{eqnarray}
which we propose to use as ``best guesses'' when setting up the deformation
for a heliospheric field model. To our knowledge, this is the first time that
simulations have been used to quantify the magnetic field-induced flattening
and deflection of the heliotail.

While the cross-sectional flattening shows the expected behavior, i.e.\ it
increases with both $B_{\rm ism}$ and distance $-z$, the fact that the derived
displacement does not depend monotonously on $z$ is somewhat at odds with the
notion of the heliotail axis being deflected by a fixed angle against the
inflow axis (which would imply $m_{\rm fit} \sim z$ for fixed $B_{\rm ism}$).
At this point, we may only note that while the simulations at hand do not
support a corresponding scaling relation, the absolute magnitude of
deflection is indeed rather small, in agreement with \citet{Wood_EA-2014},
who present observational constraints to argue for a deflection angle of the
order of $10^{\circ}$ and certainly below $20^{\circ}$.

The resulting distribution of $m_{\perp, {\rm fit}}(B_{\rm ism},z)$ of absolute
size $(11 \pm 6)$~AU and zero mean did not show any discernible systematics,
and is therefore most likely due to the probabilistic nature of the
hybrid MHD--kinetic approach (which can only accommodate a finite number of
computational particles for the neutral component). Besides this, the only
other cause for a departure of the problem's symmetry with respect to the
${\bf u}$--${\bf B}$ plane at infinity would be the influence of the solar
magnetic field, which, however, is apparently too weak to cause a systematic
asymmetry of noticeable magnitude, in contrast to what has recently been
claimed by \citet{Opher_EA-2015}. Therefore, the perpendicular displacement
$m_{\perp}$ has not been considered further after fitting was completed. \\

\subsection{Cross-sectional Variations}
\label{sec:fit-csvari}

\begin{figure}[b]
    \begin{center}
      \vspace*{8mm}
      \includegraphics[width=0.47\textwidth]{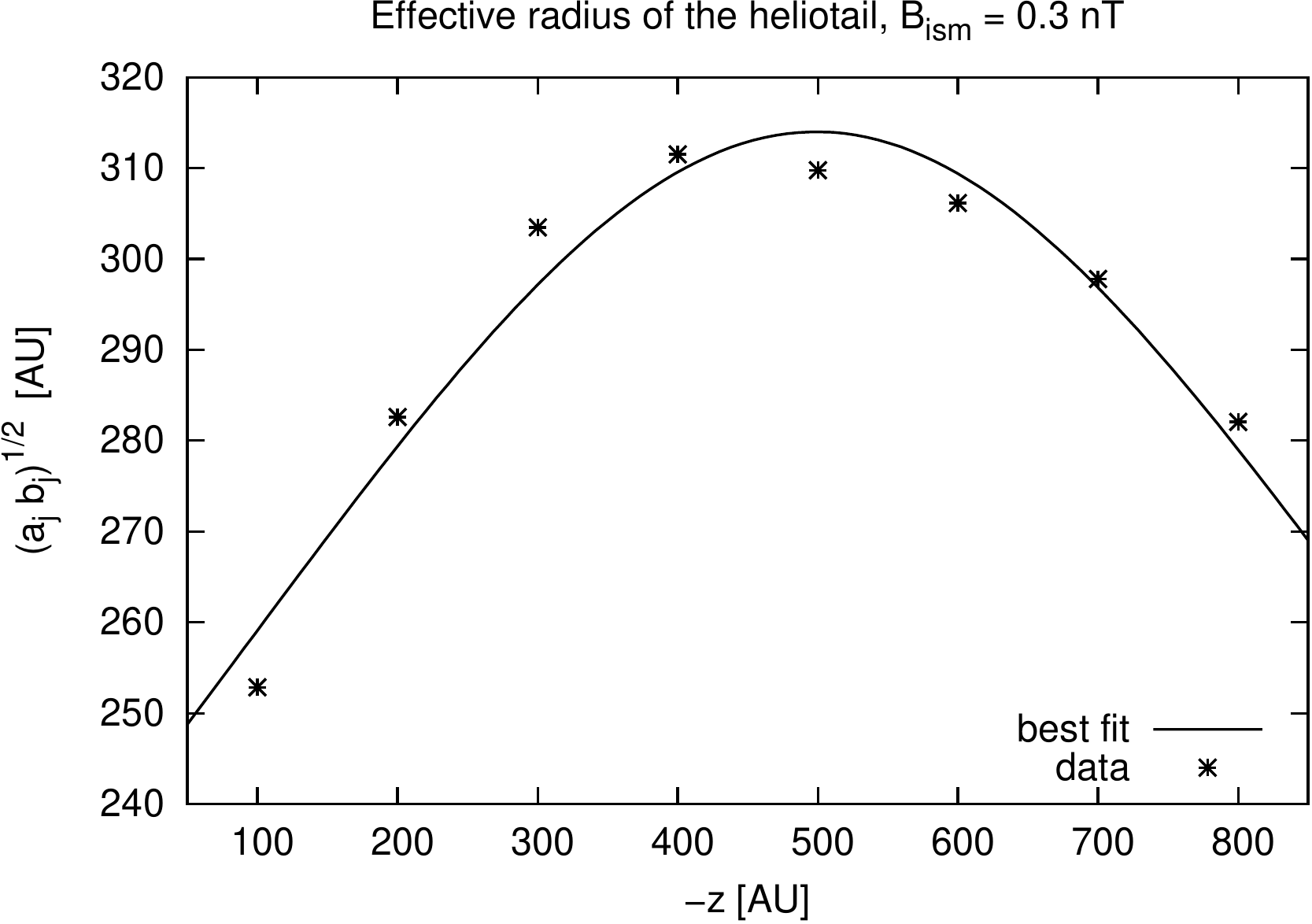}
    \end{center}
    \caption{\label{fig:fit_rad}
      Variation in the heliotail's effective cross-sectional radius for the
      0.3~nT case vs.\ fitting function (\ref{eq:fit-rad}). \\
    }
  \end{figure}

Since the distortion flow is to be incompressible, the constancy of
the product $a(z) \, b(z)$ is a necessary criterion for the consistency of
the method in its most basic form as presented in
Section~\ref{sec:deriv-const}. It is therefore interesting to see to what
degree the cross sections of our fitting ellipses can be considered constant.
Fig.~\ref{fig:fit_rad} shows the effective cross-sectional radius
$R_j := \sqrt{a_j b_j}$ for the 0.3~nT case.
%
 The latter apparently increases
toward a maximum at $z\approx -500$~AU and then decreases noticeably
toward the far tail. This behavior may at least partly be attributed to the
fact that the exact heliopause location is somewhat smeared in the
simulations, especially in the distant tail, but also to variations of $u_z$
and/or density along $z$, which are not captured in the simpler form of this
approach of constant $a \, b$. Therefore, we proceed to apply the more
general deformation as outlined in Section~\ref{sec:deriv-noco}, thereby
relaxing the requirement of cross-sectional constancy.

To this end, we first note that according to Fig.~\ref{fig:fit_rad}, the
effective radius for the parameter range in question may be reasonably
approximated by the heuristic function
\begin{equation}
  R_{\rm fit}(z) := R_{\rm c}
  \left[ 1+ \left(\frac{z-z_{\rm c}}{\delta_{\rm c}}\right)^2 \right]^{-1/2}
  \label{eq:fit-rad}
\end{equation}
with best-fit parameters
$(R_{\rm c}, z_{\rm c}, \delta_{\rm c})= (314, -499, 583)$ AU.
This varying radius is obviously related to the dimensionless squeezing
functions $a(z)$ and $b(z)$ via
\begin{equation}
  r_{\rm fit}(z) := \frac{R_{\rm fit}(z)}{R_{\rm fit}(0)} =
  \sqrt{ a(z) \, b(z) } \ .
\end{equation}
As can be seen from Fig.~\ref{fig:helli_a-r}, the cross sections can be
extrapolated to an approximately circular shape at the reference height
$z=0$, in agreement with \mbox{$a(0)=b(0)=1$}. This also means that it is
indeed consistent to place the Sun in the $z=0$ plane.
From Eq.~(\ref{eq:zzp}), we thus get
\begin{equation}
  [R_{\rm fit}(0)]^2 \ z_0 = \int_0^z [R_{\rm fit}(\tilde{z})]^2 \ \dd \tilde{z}
\end{equation}
and hence
\begin{eqnarray}
  \frac{z_0}{1+(z_{\rm c}/\delta_{\rm c})^2} &=& \delta_{\rm c} \arctan
  \left. \left( \frac{\tilde{z} - z_{\rm c}}{\delta_{\rm c}} \right)
  \right|_0^z \\
  \Leftrightarrow \ z_0 &=& C_1 \left[ \arctan
    \left( \frac{z-z_{\rm c}}{\delta_{\rm c}} \right) + C_0 \right]
\end{eqnarray}
with constants
\begin{eqnarray}
  C_0 &:=& \arctan \left( z_{\rm c} / \delta_{\rm c} \right) \\
  C_1 &:=& \delta_{\rm c} + z_{\rm c}^2/\delta_{\rm c}
\end{eqnarray}
which in this particular case evaluate to $C_0=0.707$ and $C_1=1{,}010$~AU.
If need be, this can immediately be inverted, yielding the explicit form of
$z=K(z_0)$ as
\begin{equation}
  \label{eq:zzp-helio}
  z = z_{\rm c} + \delta_{\rm c} \tan \left( \frac{z_0}{C_1}-C_0 \right) \ .
\end{equation}
The two squeezing functions then become
\begin{eqnarray}
  \label{eq:squeeze-a}
  a(z) &=& \sqrt{(a \, b) \frac{a}{b}}
  = r_{\rm fit}(z) \, \sqrt{\eta_{\rm fit}(0.3 \ {\rm nT}, z)} \\
  \label{eq:squeeze-b}
  b(z) &=& \sqrt{(a \, b) \frac{b}{a}}
  = r_{\rm fit}(z) \, \frac{1}{\sqrt{\eta_{\rm fit}(0.3 \ {\rm nT}, z)}}
\end{eqnarray}
such that all of the required information is available to turn any
(heliospheric or other) magnetic field model with circularly constant cross
sections into a more realistic version of itself by adjusting both the aspect
ratio and the absolute cross-sectional area to a desired profile. \\

\subsection{Application to the Local ISMF}
\label{sec:app_helio}

As an illustrative application, we choose the recent analytical ${\bf B}$
field model of the outer heliosheath by \citet{Roeken_EA-2015}, which
provides the exact ISMF solution to the stationary induction equation
(\ref{eq:induct}) for the case that the flow ${\bf u}$ derives from the
Rankine half-body potential $\Phi = u_0 (q/r+z)$ via
${\bf u} = -\nabla \Phi$. Here, $u_0$ and $4\pi u_0 q$,
respectively, denote the LISM flow velocity at infinity and the SW source
strength, and $r := \sqrt{x^2+y^2+z^2}$ is the Sun-centered radial
distance. The quantity $\sqrt{q}$ may be interpreted as the upstream
stand-off distance to the stagnation point, such that $q=(125~{\rm AU})^2$
may be considered a reasonable choice for the solar case.
The solution shows the expected behavior of the magnetic field piling up in
front of and draping around the heliopause, which is easily identified as
the surface satisfying
\begin{equation}
  \label{eq:hp-def}
  H({\bf r}) := 2q-(x^2+y^2) - z \sqrt{4q - (x^2+y^2)} = 0 \ .
\end{equation}
This model thus fulfills the criteria of
(i) inclusion of both ${\bf B}$ and ${\bf u}$ fields which
(ii) satisfy Eqs.~(\ref{eq:divb-0}) and (\ref{eq:induct}), and 
(iii) feature a heliopause with circular cross section,
and may therefore be subjected to the procedure outlined in the previous
subsection. The distorted field ${\bf B}$ at position ${\bf r}$ is thus
obtained by the following sequence of steps.
\begin{enumerate}
\item Find the corresponding position ${\bf r}_0$ in undistorted space
  via Eqs.~(\ref{eq:xxpyyp}) and (\ref{eq:zzp-helio}), in which $a(z)$,
  $b(z)$, and $m(z)$ are fixed using
  Eqs.~(\ref{eq:squeeze-a})--(\ref{eq:squeeze-b}) and
  (\ref{eq:fit-eta})--(\ref{eq:fit-m}).
  Since the upwind half-space $z \ge 0$ is taken to be undistorted (implying
  $\eta(z)|_{z \ge 0}=1$), the distorting transformation has to be limited to
  the tail  region, i.e.\ the downwind half-space. Moreover, since the
  transformation formula (\ref{eq:distort}) involves spatial gradients in
  $a(z)$ and $b(z)$ that would then lead to discontinuous B field components
  at $z=0$, we introduce a spatial averaging function
  \begin{equation}
    f_{\rm avg}(z) := \frac{1}{2} \bigg[ 1+
    \tanh \left(\frac{z}{80~{\rm AU}} \right) \bigg]
  \end{equation}
  and replace all four functions
  $\mu(z) \in \{ a(z), b(z), $ $m(z), F(z) \}$ by
  \begin{equation}
    \tilde{\mu}(z) := \mu(z) \ [1-f_{\rm avg}(z)] + \mu_0 \ f_{\rm avg}(z) \ ,
  \end{equation}
  thereby allowing them to tend smoothly to their undisturbed values
  $(a_0, b_0, m_0, F_0)=(1,1,0,z_0)$, and ensuring that the distorted fields
  remain continuous (and even differentiable) due to the now finite
  derivatives in transformation (\ref{eq:distort}).
\item The undistorted components of ${\bf B}_0$ are evaluated at this
  position, in this case using Eqs.~(57)--(59) of \citet{Roeken_EA-2015}.
\item The distorted components are computed from the undistorted ones using
  Eq.~(\ref{eq:distort}) in conjunction with the above parameters.
  Remember that while $m\p(z)=0$ according to Eq.~(\ref{eq:fit-m}),
  the required derivatives $a\p(z)$ and $b\p(z)$ attain a more complicated
  form, which, however, may be obtained straightforwardly from their
  definitions.
\end{enumerate}

\begin{widetext}
  \begin{center}
    \begin{figure}[t]
      \includegraphics[width=\textwidth]{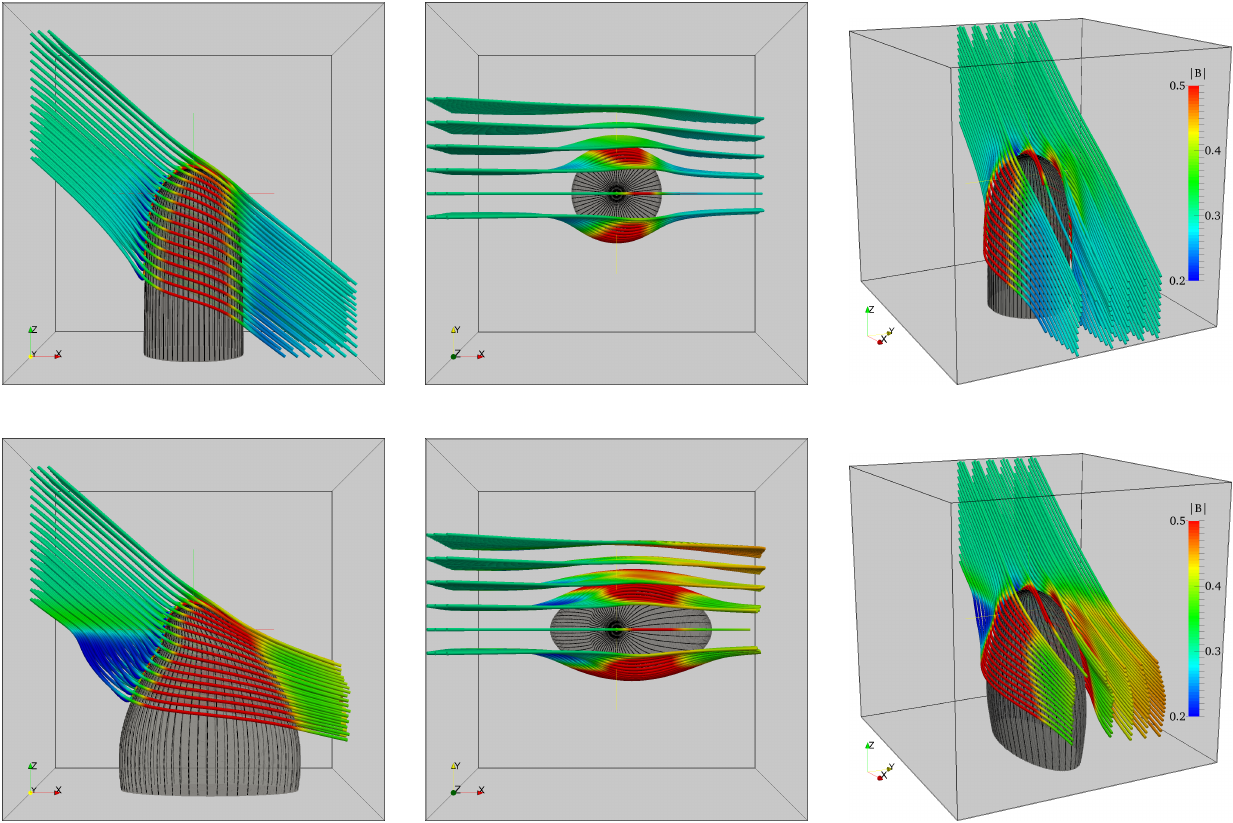}
      \caption{\label{fig:rkf-distort}
        Top row: perspective rendering of selected ${\bf B}_0({\bf r}_0)$
        field lines of the outer heliosheath model by \citet{Roeken_EA-2015}
        for $B_{\rm ism}=0.3$~nT, as seen from three different vantage points.
        The gray surface is the heliopause as defined via $H({\bf r}_0)=0$
        according to Eq.~(\ref{eq:hp-def}), to which the magnetic field is
        tangential. The black lines on the heliopause are streamlines
        emanating from the stagnation point.
        Bottom row: the same situation for ${\bf B}({\bf r})$ and
        $H({\bf r})=0$, i.e.\ after the distortion has been applied to
        both the magnetic field and the heliopause surface. Note that
        despite the massive change in shape, the magnetic field remains
        tangential to the heliopause. $\|{\bf B}\|$ values above 0.5~nT are
        capped.
      }
    \end{figure}
  \end{center}
\end{widetext}
The resulting configuration can be seen in the side-by-side comparison of
Fig.~\ref{fig:rkf-distort}. Evidently, the topological properties of the
field have been maintained, while the shape of the heliopause has been
deformed considerably.

\subsection{On the Possibility for Further Extensions}

In the deformations employed so far, the squeezing functions $a$ and $b$ have
been allowed to vary in $z$ but not within a given $x$--$y$ plane.
Therefore, by construction, they cannot accommodate asymmetric
cross-sectional shapes which are found, e.g.\ in some of the simulation runs
presented by \citet{Wood_EA-2014}. The method of distortion flows can,
however, be easily generalized to such cases. To demonstrate this potential,
we introduce a second parameter $\beta>0$ and consider the distortion flow
field (\ref{eq:w0}) to be replaced by the slightly more complicated field
\begin{equation}
  \label{eq:w1}
  {\bf w}_1: \
  \colvec{c}{ x \\ y } \mapsto \alpha
  \left[ \colvec{r}{ x \\ -y} + \beta \colvec{c}{ x^2 \\ -2 x y} \right]
\end{equation}
mapping $\mathbb{R}^2 \rightarrow \mathbb{R}^2$, in which the $y$ component
of the $\beta$ term introduces a deformation that varies linearly in $x$
(changing sign at the former symmetry plane $x=0$), and the $x$ component of
that term is chosen such that \mbox{$\nabla \cdot {\bf w}_1=0$} is satisfied.
The corresponding, now coupled set of equations of motion
$\dot{\bf r}(t) = {\bf w}_1[{\bf r}(t)]$ has the analytical solution
\begin{eqnarray}
  x(t) &=& x_0 \big[ (x_0 \beta+1) \exp(-\alpha t)-x_0 \beta \big]^{-1} \\
  y(t) &=& y_0 \exp(-\alpha t) \big[ x_0 \beta \
  [\exp(\alpha t)-1]-1 \big]^2 \ .
\end{eqnarray}
Evaluating these expressions at $t=t_1$, and defining a new constant
\begin{equation}
  c := (1/b-1) \beta
\end{equation}
with $b = \exp(-\alpha t_1)$ as before, we obtain the transformation
\begin{eqnarray}
  x &=& (x_0 /  b) (1 - c \, x_0)^{-1} \\
  y &=& (y_0 \, b) (1 - c \, x_0)^2 \ ,
\end{eqnarray}
from which the geometric properties of the distorted contour may be easily
derived.

\begin{figure}[h]
  \begin{center}
    \includegraphics[width=0.48\textwidth]{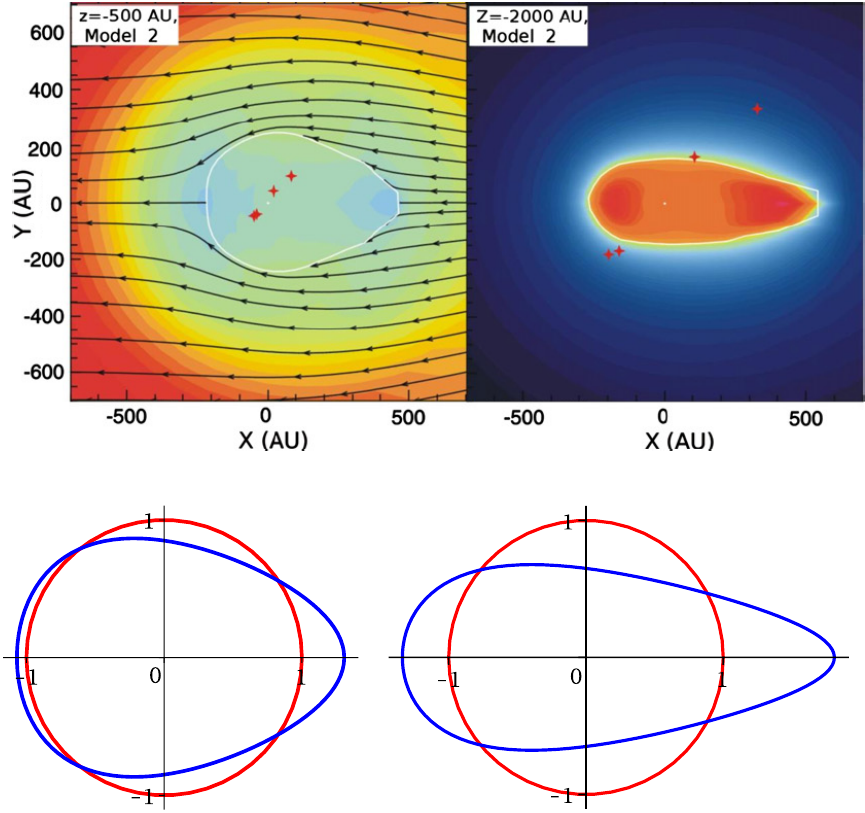} 
  \end{center}
  \caption{\label{fig:wood_comp}
    Upper panel: cross-sectional shapes of the heliopause at $z=-500$~AU
    (left) and $z=-2000$~AU (right), as seen in the simulations by
    \citet{Wood_EA-2014}, adapted from Fig.~5 of that work.
    Lower panel: two distorted contours (blue) as approximations to the
    simulated cross sections, plotted relative to the reference circle
    corresponding to $b=1$ and $c=0$ (red), which has the same total area
    due to $\nabla \cdot {\bf w}_1=0$. The respective parameters employed
    for the left and right contour are $(b=0.85, c=0.10)$ and
    $(b=0.60, c=0.15)$. \\
  }
\end{figure}

Note, in particular, that while $b$ may still be interpreted as the contour's
absolute extension along the $y$ axis in both directions, the position of
maximum extension is shifted away from $(x,y)=(0,\pm b)$ toward positive $x$,
while in the perpendicular direction, the contour extends to
$\left( [(c \pm 1) \ b]^{-1}, 0 \right)$ rather than $(\pm b^{-1},0)$. This
also implies that a spatially bounded and simply connected contour requires
$c < 1$. Hence, $b$ continues to adjust the scaling $\propto (1/b,b)$, while
the parameter $c$ controls the degree of deformation, ranging from the usual
ellipse ($c=0$) to something like a rounded triangle for values near unity. \\
After some algebra, the resulting vector transformation turns out to be
\begin{eqnarray}
  P_x &=& \frac{(1+b c \, x)^2}{b} \, P_{0x} +
  \frac{(b^2 c\p \, x - b\p) \, x}{b} \, P_{0z} \\
  P_y &=& 2 c \, (1+b c \, x) y \, P_{0x}
  + \frac{(1+2 b c \, x) b}{(1+b c \, x)^2} \, P_{0y}  \\
  &&  \nonumber + \big[ 1 + (b c + b\p c + 2 b c\p) x \big] y \, P_{0z} \\
  P_z &=& P_{0z}
\end{eqnarray}
since $z=z_0$ when embedding the 2D flow field (\ref{eq:w1}) into 3D space.
The proof of explicit compliance with constraints (\ref{eq:divb-0}) and
(\ref{eq:induct}) may be carried out in close analogy to the one presented in
Appendix~\ref{app:cunning_proof}, and is not given here.
Fig.~\ref{fig:wood_comp} illustrates how the parameters $b$ and $c$
may be used to approximate an asymmetric heliopause. \\

\section{Summary and Conclusions}
\label{sec:summary}

We have introduced and described the mathematical procedure of distortion
flows, by which any pair of 3D magnetic and velocity fields ${\bf B}$ and
${\bf u}$ can be deformed into different geometric shapes --- but with the
same topological properties --- that are more desirable in a given
application. It was proven rigorously that the method conserves both
\mbox{$\nabla \times ({\bf u} \times {\bf B})={\bf 0}$} and
\mbox{$\nabla \cdot {\bf B}=0$}.

For our principal application, namely, the heliospheric tail, we employed
sophisticated global simulations to derive heuristic profiles for the
former's cross-sectional aspect ratio and deflection as functions of position
along the tail axis, and used this information to deform a recent analytic
tail model into a still analytic, yet more realistic version of itself using
the distortion flow method.

It should, however, be kept in mind that vector equations different from
Eqs.~(\ref{eq:divb-0}) and (\ref{eq:induct}) are not guaranteed to remain
unaffected by the squeezing transformation. For instance, the
\citet{Schwadron_EA-2014} heliosphere model is built around the requirement
that there be no currents outside the heliopause
($\nabla \times {\bf B}={\bf 0}$), and while our approach could certainly
be used to deform their magnetic field, the resulting ${\bf B}$ field
will most likely have $\nabla \times {\bf B} \ne {\bf 0}$.

We note that the use of a (solenoidal) distortion flow does not imply any
constraints on the solenoidality (or lack thereof) of the physical fields
that are being distorted. In particular, we do not in any way argue for the
plasma to be incompressible.

Furthermore, we would like to stress that the applicability of the
distortion flow approach outlined in this paper is suitable for any source
magnetic field, and does not depend on its particular physical context.
Therefore, it is a rather generic tool that could clearly be applied outside
the heliospheric context as well. In view of the rather small number of known
analytic solutions to the MHD induction equation, the method clearly has the
potential to yield the solutions for the magnetized flow around complex
geometries simply by starting from a known solution of simpler geometry
(like a sphere, see, e.g.\ the appendix of \citet{Isenberg_EA-2015} and
references therein) and then deforming this solution as desired. \\

\acknowledgments
\section*{Acknowledgments}

We are grateful to Gunnar Hornig, Yuri Litvinenko, Frederic Effenberger, and
Andreas Kopp for helpful discussions.\ We acknowledge financial support via
the project FI~706/15-1 funded by the Deutsche Forschungsgemeinschaft (DFG).
J.H. acknowledges support from NASA grants NNX12AB30G, NNX14AF43G,
NNX14AF43G, NSF grant OCI-1144120, and Department of Energy grant
SC0008334.
We also appreciate discussions at the workshop ``Reconnection, Turbulence,
and Particles in the Heliosphere'' in Queenstown, New Zealand, as well as
at the team meeting ``Heliosheath Processes and Structure of the Heliopause:
Modeling Energetic Particles, Cosmic Rays, and Magnetic Fields'' supported
by the International Space Science Institute (ISSI) in Bern, Switzerland.

\appendix
\numberwithin{equation}{section}

\section{A. Invariance of Solenoidality and Frozen-in
  Conditions}
\label{app:cunning_proof}

We proceed to prove that if the vector fields ${\bf u}$ and ${\bf B}$
are computed using transformation (\ref{eq:distort}) from fields ${\bf u}_0$
and ${\bf B}_0$ satisfying $\nabla_0 \cdot {\bf B}_0=0$ and
$\nabla_0 \times ({\bf u}_0 \times {\bf B}_0) = {\bf 0}$, these two equations
in deformed coordinates will also hold for ${\bf u}$ and ${\bf B}$.
To this end, we first need to see how the differential operator
$\nabla=(\partial_x, \partial_y, \partial_z)$ in distorted space is related
to its undistorted counterpart
$\nabla_0=(\partial_{x_0}, \partial_{y_0}, \partial_{z_0})$.
From Eqs.~(\ref{eq:xxpyyp}) and (\ref{eq:zzp}), we have
\begin{equation}
  \colvec{c}{\partial_x \\ \partial_y \\ \partial_z } =
  \colvec{ccc}{ 1/a & 0 & 0 \\ 0 & 1/b & 0 \\
    -(a\p x + m\p)/a & -(y \ b\p) /b & a b }
  \colvec{c}{\partial_{x_0} \\ \partial_{y_0} \\ \partial_{z_0} } \ ,
\end{equation}
which is found by explicit calculation via
\begin{eqnarray}
  \frac{\partial}{\partial x}
  &=& \frac{\partial x_0}{\partial x} \frac{\partial}{\partial x_0}
  +   \frac{\partial y_0}{\partial x} \frac{\partial}{\partial y_0}
  +   \frac{\partial z_0}{\partial x} \frac{\partial}{\partial z_0}
  =  \frac{\partial}{\partial x}
  \left( \frac{x - m(z)}{a(z)} \right) \frac{\partial}{\partial x_0}
  + 0 + 0 = \frac{1}{a(z)} \frac{\partial}{\partial x_0} \\
  \frac{\partial}{\partial y}
  &=& \frac{\partial x_0}{\partial y} \frac{\partial}{\partial x_0}
  +   \frac{\partial y_0}{\partial y} \frac{\partial}{\partial y_0}
  +   \frac{\partial z_0}{\partial y} \frac{\partial}{\partial z_0}
  = 0 + \frac{\partial}{\partial y}
  \left( \frac{y}{b(z)} \right) \frac{\partial}{\partial y_0}
  + 0 = \frac{1}{b(z)} \frac{\partial}{\partial y_0} \\
  \frac{\partial}{\partial z}
  &=& \nonumber
  \frac{\partial x_0}{\partial z} \frac{\partial}{\partial x_0}
  + \frac{\partial y_0}{\partial z} \frac{\partial}{\partial y_0}
  + \frac{\partial z_0}{\partial z} \frac{\partial}{\partial z_0} \\
  &=& \nonumber
  \frac{\partial }{\partial z}
  \left( \frac{x - m(z)}{a(z)} \right) \frac{\partial}{\partial x_0}
  + \frac{\partial}{\partial z}
  \left( \frac{y}{b(z)} \right) \frac{\partial}{\partial y_0}
  + \frac{\partial F(z)}{\partial z} \frac{\partial}{\partial z_0} \\
  &=& - \frac{(x - m(z)) \, a\p(z) + m\p(z) \, a(z)}{a^2(z)}
  \frac{\partial}{\partial x_0}
  - \frac{y \, b\p(z)}{b^2(z)} \frac{\partial}{\partial y_0}
  + a(z) \, b(z) \frac{\partial}{\partial z_0} \\
  &=& \nonumber 
  - \frac{x \, a\p(z) + m\p(z)}{a(z)}
  \frac{\partial}{\partial x_0}
  - \frac{y \, b\p(z)}{b(z)} \frac{\partial}{\partial y_0}
  + a(z) \, b(z) \frac{\partial}{\partial z_0} \ .
\end{eqnarray}
Then, by means of transformation (\ref{eq:distort}), we obtain
\begin{equation}
  \begin{split}
    \nabla \cdot {\bf P}
    =&\
    \frac{\partial P_x}{\partial x} +
    \frac{\partial P_y}{\partial y} +
    \frac{\partial P_z}{\partial z} \\
    =&\
    \left( \frac{1}{a} \frac{\partial}{\partial x_0} \right)
    \left( a \, P_{0x} + \frac{a\p \, x_0 + m\p}{a \, b} \, P_{0z} \right) +
    \left( \frac{1}{b} \frac{\partial}{\partial y_0} \right)
    \left( b \, P_{0y} + \frac{b\p \, y_0}{a \, b} \, P_{0z} \right) \\
    &+ \left(
      -\frac{a\p \, x_0 + m\p}{a} \frac{\partial}{\partial x_0}
      -\frac{b\p \, y_0      }{b} \frac{\partial}{\partial y_0}
      + a \, b \, \frac{\partial}{\partial z_0} \right) \frac{P_{0z}}{a \, b}
    \\
    =&\
    \frac{\partial P_{0x}}{\partial x_0} + \frac{1}{a \, b} \left[
      \frac{\partial}{\partial x_0}
      \left(\frac{a\p \, x_0 + m\p}{a} \, P_{0z} \right) +
      \frac{\partial}{\partial y_0}
      \left(\frac{b\p \, y_0      }{b} \, P_{0z} \right)
    \right] + \frac{\partial P_{0y}}{\partial y_0} \\
    &- \frac{1}{a \, b} \left[ \left(
        \frac{a\p \, x_0 + m\p}{a} \frac{\partial P_{0z}}{\partial x_0}
      \right) + \left(
        \frac{b\p \, y_0    }{b} \frac{\partial P_{0z}}{\partial y_0}
      \right) \right] + \frac{\partial P_{0z}}{\partial z_0}
    + P_{0z} \, a \, b \ \frac{\partial}{\partial z_0}
    \left( \frac{1}{a \, b} \right) \\
    =&\ \nabla_0 \cdot {\bf P}_0 + \frac{P_{0z}}{a \, b} \bigg[
    \underbrace{\frac{\partial}{\partial x_0}
      \left( \frac{a\p \, x_0 + m\p}{a} \right)}_{= a\p/a} +
    \underbrace{\frac{\partial}{\partial y_0}
      \left( \frac{b\p \, y_0  }{b} \right)}_{= b\p/b} \bigg] - P_{0z} \left[
      \frac{1}{a} \frac{\partial a}{\partial z_0} +
      \frac{1}{b} \frac{\partial b}{\partial z_0} \right] \ ,
  \end{split}
\end{equation}
in which the final pair of square brackets cancels the preceding one due to
\begin{equation}
  a\p = \frac{\partial a(z)}{\partial z}
  = a \, b \ \frac{\partial a}{\partial z_0}
\end{equation}
and similarly for $b\p = \partial_{z} b(z)$. This proves the invariance of
the solenoidality condition for ${\bf P}$. \\

For the induction equation, we start by defining the undistorted electric
field ${\bf E}_0 := - {\bf u}_0 \times {\bf B}_0$, such that
$\nabla_0 \times {\bf E}_0 = {\bf 0}$ according to Eq.~(\ref{eq:induct}),
and then evaluate
\begin{equation}
  \begin{split}
    - {\bf E} &= {\bf u} \times {\bf B} \\
    &= \colvec{c}{
      a \, u_{0x} + (a\p x_0 + m\p)/(a b) \ u_{0z} \\
      b \, u_{0y} + (b\p y_0      )/(a b) \ u_{0z} \\ u_{0z}/(a b) } \times
    \colvec{c}{
      a \ B_{0x} + (a\p x_0 + m\p)/(a b) \ B_{0z} \\
      b \ B_{0y} + (b\p y_0      )/(a b) \ B_{0z} \\ B_{0z}/(a b) } \\
    &= \colvec{c}{
      (u_{0y} B_{0z} - u_{0z} B_{0y}) / a \\
      (u_{0z} B_{0x} - u_{0x} B_{0z}) / b \\
      (u_{0x} B_{0y} - u_{0y} B_{0x}) \, a b
      - (u_{0y} B_{0z} - u_{0z} B_{0y}) (a\p x_0 + m\p)/a
      - (u_{0z} B_{0x} - u_{0x} B_{0z}) \ b\p y_0 / b } \\
    &= -\colvec{c}{ E_{0x} / a \\ E_{0y} / b \\
      E_{0z} \, a \, b - E_{0x} \, (a\p x_0 + m\p)/a - E_{0y} \, b\p y_0 / b }
  \end{split}
\end{equation}
and further
\begin{equation}
  \nabla \times {\bf E} =
  \colvec{c}{ (1/a) \ \partial_{x_0} \\ (1/b) \ \partial_{y_0} \\
    -(a\p x_0 + m\p)/a \ \partial_{x_0}
    -(b\p y_0      )/b \ \partial_{y_0}
    +        a \, b    \ \partial_{z_0} } \times
  \colvec{c}{ E_{0x} / a \\ E_{0y} / b \\
    E_{0z} \ a \, b - E_{0x} \, (a\p x_0 + m\p)/a - E_{0y} \ b\p y_0 / b } \ .
\end{equation}
The $x$ component reads
\begin{equation}
  \begin{split}
    (\nabla \times {\bf E})_x =&\
    \frac{1}{b} \, \frac{\partial}{\partial y_0}
    \left( E_{0z} \, a b - E_{0x} \, \frac{a\p x_0 + m\p}{a}
      - E_{0y} \, \frac{b\p y_0}{b} \right)
    - \left( - \frac{a\p x_0 + m\p}{a} \, \frac{\partial}{\partial x_0}
      -\frac{b\p y_0      }{b} \, \frac{\partial}{\partial y_0}
      +  a b \, \frac{\partial}{\partial z_0} \right) \frac{E_{0y}}{b} \\
    =&\ \bigg(
    \underbrace{\frac{\partial E_{0z}}{\partial y_0}
      - \frac{\partial E_{0y}}{\partial z_0}}_{= (\nabla_0 \times {\bf E}_0)_x}
    \bigg) a +
    \bigg( \underbrace{\frac{\partial E_{0y}}{\partial x_0}
      - \frac{\partial E_{0x}}{\partial y_0}}_{= (\nabla_0 \times {\bf E}_0)_z}
    \bigg) \frac{a\p x_0 + m\p}{a \, b}
    + \left( \frac{a}{b} \frac{\partial b}{\partial z_0} - \frac{b\p}{b^2}
    \right) \ E_{0y} = 0 \ ,
  \end{split}
\end{equation}
in which the last term vanishes for the same reason as above.
The  $y$ component is computed in complete analogy, and vanishes as well.
Finally,
\begin{equation}
  (\nabla \times {\bf E})_z =
  \frac{1}{a} \frac{\partial}{\partial x_0} \left(\frac{E_{0y}}{b} \right) -
  \frac{1}{b} \frac{\partial}{\partial y_0} \left(\frac{E_{0x}}{a} \right) 
  = \frac{1}{a \, b}
  \bigg( \underbrace{\frac{\partial E_{0y}}{\partial x_0} -
    \frac{\partial E_{0x}}{\partial y_0}}_{= (\nabla_0 \times {\bf E}_0)_z}
  \bigg) = 0 \ ,
\end{equation}
which concludes the proof of the invariance of the induction equation under
transformation (\ref{eq:distort}). \\

The above considerations can be extended to more general distortion flow fields
${\bf w}$ as follows. Since in this case the transition from ${\bf P}_0$ to
${\bf P}$ proceeds via the evolution equation (\ref{eq:dPdt_distort}),
it is convenient for the proof of the invariance of constraint
(\ref{eq:divb-0}) to consider
\begin{equation}
  \begin{split}
  \partial_t (\nabla \cdot {\bf P})
  = \nabla \cdot (\partial_t{\bf P})
  &= \nabla \cdot \big[ \nabla \times ({\bf w} \times {\bf P})
  - (\nabla \cdot {\bf P}) {\bf w}
  + (\nabla \cdot {\bf w}) {\bf P} \big] \\
  &= 0 - \big[ (\nabla \cdot {\bf P}) (\nabla \cdot {\bf w})
  + {\bf w} \cdot \nabla (\nabla \cdot {\bf P}) \big]
  + \big[ (\nabla \cdot {\bf w}) (\nabla \cdot {\bf P})
  + {\bf P} \cdot \nabla (\nabla \cdot {\bf w}) \big] \\
  &=
  {\bf P} \cdot \nabla (\nabla \cdot {\bf w}) -
  {\bf w} \cdot \nabla (\nabla \cdot {\bf P}) \ .
\end{split}
\end{equation}
If $\nabla \cdot {\bf w}=0$, then this is the well-known advection partial
differential equation (PDE)
\begin{equation}
  (\partial_t + {\bf w} \cdot \nabla) f = 0
\end{equation}
for $f({\bf r},t):=\nabla \cdot {\bf P}$, meaning that $f$ is passively
advected (and thus constant) along flow lines of ${\bf w}$. In particular,
$f({\bf r},t) = 0$ is the only solution that satisfies the appropriate
initial condition $f({\bf r},0) = (\nabla \cdot {\bf P})|_{t=0} = 0$ for all
${\bf r}$. This shows explicitly that $\nabla \cdot {\bf P}$ continues to
vanish as long as $\nabla \cdot {\bf w}=0$. In fact, we see that any
${\bf w}$ with constant (not necessarily vanishing) divergence, like for
instance ${\bf w} \propto {\bf r}$, will conserve the solenoidality of
${\bf P}$. This is intuitively plausible, since such a linear scaling
corresponds to an isotropic ``magnification'' of space that leaves all angles
between vectors, etc.\ unchanged.

Regarding the second constraint (\ref{eq:induct}), we first define the vector
quantity to be conserved as
${\bf C} := \nabla \times ( {\bf u} \times {\bf B} )$, for which we need to
show that for the initial condition ${\bf C}|_{t=0} = {\bf 0}$,
\begin{equation}
  \label{eq:c-vec1}
  \begin{split}
 \partial_t  {\bf C} =& \ \partial_t
    \big( \nabla \times ( {\bf u} \times {\bf B} ) \big)
    = \nabla \times \big( [\partial_t {\bf u}] \times {\bf B}
    + {\bf u} \times [\partial_t {\bf B}] \big) \\
    =& \ \nabla \times \big( [ ({\bf u} \cdot \nabla) {\bf w}
    - ({\bf w} \cdot \nabla) {\bf u} ]
    \times {\bf B} + {\bf u} \times [
    ({\bf B} \cdot \nabla) {\bf w}
    - ({\bf w} \cdot \nabla) {\bf B}
    ] \big)
  \end{split}
\end{equation}
vanishes at any $t>0$. Using the identities
\begin{eqnarray}
  \big[ ( {\bf A} \cdot \nabla ) {\bf B} \big]_i &=& \ A_j \partial_j B_i \\
  \big[ \nabla \times ({\bf A} \times {\bf B}) \big]_i &=& \
  \eps_{ijk} \, \partial_j ( \eps_{klm} A_l B_m)
  = (\delta_{il} \delta_{jm} - \delta_{im} \delta_{jl})
  \, \partial_j ( A_l B_m) = \partial_j ( A_i B_j - A_j B_i)
\end{eqnarray}
for general vectors ${\bf A}, {\bf B} \in \mathbb{R}^3$ (with the convention
$\varepsilon_{123}=1$, and summation over double indices is implied), we
obtain for the $i$th component of Eq.~(\ref{eq:c-vec1})
\begin{equation}
  \label{eq:c-icom}
  \begin{split}
    \partial_t C_i =\ & \partial_j \big(
    \big[ (u_k \partial_k w_i - w_k \partial_k u_i) B_j
    -     (u_k \partial_k w_j - w_k \partial_k u_j) B_i \big] \\
    & +
    \big[ u_i (B_k \partial_k w_j - w_k \partial_k B_j)
    -     u_j (B_k \partial_k w_i - w_k \partial_k B_i) \big] \big) \\
    =\ & \partial_j \big[
    (u_k B_j - u_j B_k) (\partial_k w_i) +
    (u_i B_k - u_k B_i) (\partial_k w_j) +
    \partial_k (u_j B_i - u_i B_j) w_k \big] \\
    =\ & \underbrace{\partial_j (u_k B_j - u_j B_k)}_{[\nabla \times
      ({\bf u} \times {\bf B})]_k} (\partial_k w_i)
    + \underbrace{(u_k B_j - u_j B_k) (\partial_{jk} w_i)}_{=:S}
    + \underbrace{\partial_j (u_i B_k - u_k B_i) (\partial_k w_j)}_{=:N} \\
    & + (u_i B_k - u_k B_i)
    \underbrace{(\partial_{jk} w_j)}_{[\nabla (\nabla \cdot {\bf w})]_k}
    + \underbrace{\partial_{jk} (u_j B_i - u_i B_j)}_{-\partial_k  
      [\nabla \times ({\bf u} \times {\bf B})]_i} w_k
    + \underbrace{\partial_k (u_j B_i - u_i B_j) (\partial_j w_k)}_{-N} \ ,
  \end{split}
\end{equation}
in which the second term $S$ vanishes due to the symmetry of $\partial_{jk}$
and the summation over both indices. Upon reverting to
vector notation, Eq.~(\ref{eq:c-icom}) thus simply becomes
\begin{equation}
  \label{eq:c-vec2}
  \partial_t {\bf C} = ({\bf C} \cdot \nabla) {\bf w}
  - ({\bf w} \cdot \nabla) {\bf C}
\end{equation}
when evaluated for the special case $\nabla \cdot {\bf w}=0$ that was already
found necessary for the invariance of the solenoidality constraint.
Since ${\bf w}$
does not depend on $t$, one can easily use Eq.~(\ref{eq:c-vec2}) to show via
mathematical induction that
\begin{equation}
  \label{eq:c_ic}
  (\partial_t^n {\bf C})|_{t=0} = {\bf 0} \quad
  \forall \ n \in \mathbb{N}_0 \ .
\end{equation}
The PDE system (\ref{eq:c-vec2}), together with the initial conditions
(\ref{eq:c_ic}), states a Cauchy problem. Under the real analyticity
assumption on all coefficients of this PDE system, or rather the distortion
vector field ${\bf w}$, one can apply the Cauchy-Kovalevskaya theorem for
local existence and uniqueness. From this, it follows that there exists a
unique local analytic solution of the Cauchy problem, namely, for ${\bf C}$,
in the neighborhood of $t=0$. Consequently, the Taylor series of ${\bf C}$
at $t=0$ with the conditions (\ref{eq:c_ic}) indeed yields the trivial
solution
\begin{equation}
  {\bf C}(t) = \sum_{n = 0}^{\infty} \frac{1}{n!}
  \left. \frac{\partial^n {\bf C}}{\partial t^n} \right|_{t=0}
  t^n = {\bf 0} \ ,
\end{equation}
proving that if ${\bf C}$ is zero initially, it stays zero at all later
times. Again, a vanishing gradient of $\nabla \cdot {\bf w}$ is sufficient
to obtain this result. This is analogous to the usual MHD induction equation
for $\partial_t {\bf B}$, which evolves a pre-existing magnetic field
${\bf B}$ non-linearly but cannot do so without an initial seed field.

We note in passing that it would not be permissible to conclude already from 
${\bf C}|_{t=0}={\bf 0}$ and the fact that the linear equation
(\ref{eq:c-vec2}) admits the trivial solution ${\bf C}={\bf 0}$ that
${\bf C}$ has to vanish identically, as is sometimes done erroneously in the
literature for the usual, time-dependent MHD induction equation
\citep[e.g.][]{Kandaswamy-2000}. This can clearly be seen by the simple
example of the ordinary linear differential equation
$\dot{g}(t) = (2/t) \, g(t)$, which possesses the general solution
$g(t)=g_1 \, t^2$ that is non-zero for $t>0$ and any constant
$g_1=g(1) \in \mathbb{R} \backslash \{0 \}$, despite the fact that
$g(0)=0=\dot{g}(0)$. \\

\section{B. The Case of Arbitrary Vertical
  Displacements}
\label{app:wedge}

\noindent The PDE for the determination of the new $z$ component of the
general coordinate transformation defined by Eq.~(\ref{eq:xxpyyp}) and
$z_0 = z_0(x, y, z)$, satisfying volume conservation, can be obtained by
first computing the exterior derivatives of the original coordinates
$x_0$, $y_0$, and $z_0$, reading
\begin{equation}
  \begin{split}
    & \dd x_0 = \frac{1}{a} \left[ \dd x - \left( \frac{[x-m] a\p}{a}
        + m\p \right) \dd z \right] \\
    & \dd y_0 = \frac{1}{b} \left[ \dd y - \frac{y \, b\p}{b} \,
      \dd z \right] \\ &
    \dd z_0 = \frac{\partial z_0}{\partial x} \,
    \dd x + \frac{\partial z_0}{\partial y} \,
    \dd y + \frac{\partial z_0}{\partial z} \, \dd z \ . 
  \end{split}
\end{equation}
The corresponding volume three-form becomes
\begin{equation}
  \label{eq:volform}
  \dd x_0 \wedge \dd y_0 \wedge \dd z_0 = \frac{1}{a \, b} \left[
    \left( \frac{[x - m] a\p}{a} + m\p \right)
    \frac{\partial z_0}{\partial x}
    + \frac{y \, b\p}{b} \, \frac{\partial z_0}{\partial y}
    + \frac{\partial z_0}{\partial z} \right] \,
  \dd x \wedge \dd y \wedge \dd z \ ,
\end{equation}
in which the symbol $\wedge$ denotes the standard alternating wedge product
on the exterior algebra $\Lambda(\mathbb{R}^3)$. Requiring volume
conservation, i.e.
$\dd x_0 \wedge \dd y_0 \wedge \dd z_0 = \dd x \wedge \dd y \wedge \dd z$,
we immediately find from Eq.~(\ref{eq:volform}) that
\begin{equation}
  \label{eq:z0-pde}
  \left( \frac{[x - m] a\p}{a} + m\p \right)
  \frac{\partial z_0}{\partial x} + \frac{y \, b\p}{b} \,
  \frac{\partial z_0}{\partial y} + \frac{\partial z_0}{\partial z}
  = a \, b \ . 
\end{equation}
This PDE is the defining equation for the general component transformation
$z_0 = z_0(x, y, z)$. In the special case $z_0 = z_0(z)$ discussed in
Section~\ref{sec:deriv-noco}, we have
$\partial z_0/\partial x = 0 = \partial z_0/\partial y$, and thus from
(\ref{eq:z0-pde}) we obtain
\begin{equation}
\frac{\partial z_0}{\partial z} = a \, b \ .
\end{equation}
Integration leads to  
\begin{equation}
z_0 = \int^z_0 a(\tilde{z}) \, b(\tilde{z}) \, \dd \tilde{z} \ ,
\end{equation}
which is the component transformation used in Eq.~(34). \\

\section{C. Ellipse Fitting Procedure}
\label{app:fit-ellipse}

Let
\begin{equation}
  \varphi_{i,j} := \arctan \left(\frac{y_i-m_{\perp,j}}{x_i-m_j} \right)
\end{equation}
be the angle between the $x$ axis and the vector which points from the center
of ${\cal E}_j$ to the contour point $(x_i,y_i)_j \in {\cal H}_j$.
The respective distances from the center to that point and to the
intersection with ${\cal E}_j$ in the same direction are then given by
\begin{eqnarray}
  r_{{\cal H}_j,i} &=& \left[ (x_i-m_j)^2+(y_i-m_{\perp,j})^2 \right]^{1/2} \\
  r_{{\cal E}_j,i} &=& \left[
    \left(\frac{\cos \varphi_{i,j}}{a_j}\right)^2 +
    \left(\frac{\sin \varphi_{i,j}}{b_j}\right)^2 \right]^{-1/2} \ .
\end{eqnarray}
Using the identities
$\cos^2 \varphi = 1/(1+\tan^2 \varphi)$ and
$\sin^2 \varphi = (\tan^2 \varphi)/(1+\tan^2 \varphi)$
allows us to write the quantity to be minimized as
\begin{equation}
  \begin{split}
    \Delta_j =& \sum_{i=1}^N
    \left| (r_{{\cal H}_j,i})^2 - (r_{{\cal E}_j,i})^2 \right|
    (\delta \varphi_{i,j}/2) \\
    \propto& \sum_{i=1}^N \left[ (x_i-m_j)^2+(y_i-m_{\perp,j})^2 \right] \times
    \left| 1- \left[
        \left( \frac{x_i-m_j}{a_j}       \right)^2 +
        \left( \frac{y_i-m_{\perp,j}}{b_j} \right)^2 \right]^{-1} \right| \ .
  \end{split}
\end{equation}
The contour points $(x_i,y_i)_j$ are generated with equidistant distribution
of $\arctan(y_i/x_i)_j$. Strictly speaking, $\Delta_j$ would need
$\delta \varphi_{i,j} := \varphi_{i+1,j}-\varphi_{i,j}$ to be uniform for all
$i$ in order to actually be proportional to the total difference area between
${\cal H}_j$ and ${\cal E}_j$. However, given that there is in any case some
freedom in the details of a chosen fitting method,  this discrepancy is
certainly small enough to be safely ignored for the present purpose. \\

\vspace{0.5cm}
\bibliographystyle{apj}
\bibliography{references_distort}

\end{document}